%% file: paperSubmit.tex
\documentclass[10pt,journal,romanappendices]{IEEEtran}
\usepackage{cite,calc,url}
\usepackage[cmex10]{amsmath} % ensure not too small fonts (so Type 1 still)
\usepackage{amsfonts,amssymb,graphics,psfrag,subfigure,epsfig,graphics}
\usepackage[mathcal]{euscript} % redefine \mathcal to be \EuScript;
\usepackage{kbordermatrix}

\input{gww_defs}

\input{gww_chars}
\input{gww_rvs}

\newcommand{\bzero}{\mathbf{0}}

\newcommand{\bVr}{\bV_\mrm{r}}
\newcommand{\bVn}{\bV_\mrm{n}}

\newcommand{\obHeff}{\obH_\mrm{eff}}
\newcommand{\tbHeff}{\bHtt_\mrm{eff}}
\newcommand{\bSieff}{\bSi_\mrm{eff}}
\newcommand{\tbSieff}{\tbSi_\mrm{eff}}
\newcommand{\bHt}{\bH}  % total
 
\newcommand{\bHtt}{\tilde{\bHt}}
\newcommand{\bHe}{\bHt_\mrm{e}}
\newcommand{\bHte}{\bHtt_\mrm{e}}  % total

\newcommand{\bHr}{\bHt_\mrm{r}}
\newcommand{\bHtr}{\bHtt_\mrm{r}}

\newcommand{\Nt}{{n_\mrm{t}}}
\newcommand{\Ne}{{n_\mrm{e}}}
\newcommand{\Nr}{{n_\mrm{r}}}
\newcommand{\rvyr}{\rvy_\mrm{r}}
\newcommand{\rvye}{\rvy_\mrm{e}}
\newcommand{\rvbyr}{\rvby_\mrm{r}}
\newcommand{\rvbye}{\rvby_\mrm{e}}
\newcommand{\rvbytr}{\rvbyt_\mrm{r}}
\newcommand{\rvbyte}{\rvbyt_\mrm{e}}
\newcommand{\rvbzr}{\rvbz_\mrm{r}}
\newcommand{\rvbze}{\rvbz_\mrm{e}}
\newcommand{\rvbztr}{\rvbzt_\mrm{r}}
\newcommand{\rvbzte}{\rvbzt_\mrm{e}}
\newcommand{\bPsir}{\bPsi_\mrm{r}}
\newcommand{\bPsie}{\bPsi_\mrm{e}}
\newcommand{\bPsit}{\bPsi_\mrm{t}}
\newcommand{\bSir}{\bSi_\mrm{r}}
\newcommand{\bSie}{\bSi_\mrm{e}}
\newcommand{\rvbHr}{\rvbH_\mrm{r}}
\newcommand{\rvbHe}{\rvbH_\mrm{e}}
\newcommand{\bDr}{\bD_\mrm{r}}
\newcommand{\bDev}{\bD_\mrm{e}}
\newcommand{\bPhsv}{\bDe}  
\newcommand{\bHrsv}{\bDe}
\newcommand{\bDepar}{\bXi}
\newcommand{\depar}{\xi}
\newcommand{\Rub}{R_+(\obK_P,\obK_\bPh)}
\newcommand{\Rp}{R_+}
\newcommand{\Rm}{R_-}
\newcommand{\Rpp}{R_{++}}
\newcommand{\bLab}{\bLa_\mrm{b}}  % backward
\newcommand{\tbLab}{\tilde{\bLa}_\mrm{b}}  % backward
\newcommand{\cSr}{\cS_\mrm{r}}
\newcommand{\cSe}{\cS_\mrm{e}}
\newcommand{\cSre}{\cS_\mrm{r,e}}
\newcommand{\cSn}{\cS_\mrm{n}}
\newcommand{\tbPhr}{\tbPh}
\newcommand{\tbPhe}{\cbPh}
\newcommand{\tobTh}{{\breve{\bTh}}}

\renewcommand{\Pt}{P_\mrm{t}}

\newcounter{tempcount}

\begin{document}

\title{Secure Transmission with Multiple Antennas II:\\ 
The MIMOME Wiretap Channel}
\author{Ashish Khisti~\IEEEmembership{Member,~IEEE} 
        and Gregory W.~Wornell~\IEEEmembership{Fellow,~IEEE}%
\thanks{Manuscript received August 2008, revised June 2010.
This work was supported in part by the National Science Foundation
under Grant No.~CCF-0515109.   This work was presented in part at the
Allerton Conference on Communications, Control and Computing, Sep.\
2007. }%
\thanks{A. Khisti was with the Dept.\ Electrical Engineering and
  Computer Science, Massachusetts Institute of Technology, Cambridge,
  MA 02139.  He is now with the Dept.\ Electrical and Computer
  Engineering, University of Toronto, ON, Canada M5S 3G4.  (Email:
  khisti@comm.utoronto.ca).  G. W. Wornell is with the Dept.\
  Electrical Engineering and Computer Science, Massachusetts Institute
  of Technology, Cambridge, MA 02139. (Email: gww@mit.edu)}. }

\maketitle

%\thispagestyle{fancy}
%\pagestyle{empty}

%%%%%%%%%%%%%%%%%%%%%%%%%%%%%%%%%%%%%%%%%%%%%%%%%%%%%%%%%%%%%%%%%%%%%%%%%%%%%%%%

\begin{abstract}

The capacity of the Gaussian wiretap channel model is analyzed when
there are multiple antennas at the sender, intended receiver and
eavesdropper.  The associated channel matrices are fixed and known to
all the terminals.  A computable characterization of the secrecy
capacity is established as the saddle point solution to a minimax
problem.  The converse is based on a Sato-type argument used in other
broadcast settings, and the coding theorem is based on Gaussian
wiretap codebooks.

At high signal-to-noise ratio (SNR), the secrecy capacity is shown to
be attained by simultaneously diagonalizing the channel matrices via
the generalized singular value decomposition, and independently coding
across the resulting parallel channels.  The associated capacity is
expressed in terms of the corresponding generalized singular values.
It is shown that a semi-blind ``masked'' multi-input multi-output
(MIMO) transmission strategy that sends information along directions
in which there is gain to the intended receiver, and synthetic noise
along directions in which there is not, can be arbitrarily far from
capacity in this regime.

Necessary and sufficient conditions for the secrecy capacity to be
zero are provided, which simplify in the limit of many antennas when
the entries of the channel matrices are independent and identically
distributed.  The resulting scaling laws establish that to prevent
secure communication, the eavesdropper needs 3 times as many antennas
as the sender and intended receiver have jointly, and that the
optimimum division of antennas between sender and intended receiver is
in the ratio of $2:1$.
\end{abstract}


\section{Introduction}

\IEEEPARstart{M}{ultiple} antennas are a valuable resource in wireless
communication.  Over the last several years, there there has been
extensive activity in exploring the design, analysis, and
implementation of wireless systems with multiple antennas, emphasizing
their role in improving robustness and throughput.  In this work, we
develop aspects of the emerging role of multiple antennas in providing
communication security at the physical layer.

The wiretap channel \cite{wyner:75Wiretap} is an information-theoretic
model for physical-layer security.  In the model, there are three
terminals---a sender, an intended receiver, and an eavesdropper.  The
goal is to exploit the structure of the underlying broadcast channel
to transmit a message reliably to the intended receiver, while leaking
asymptotically no information to the eavesdropper.  A single-letter
characterization of the secrecy capacity when the underlying broadcast
channel is discrete and memoryless is developed in
\cite{csiszarKorner:78}.  An explicit solution for the scalar Gaussian
case is obtained in \cite{leung-Yan-CheongHellman:99}, where the
optimality of Gaussian codebooks is established.

In this paper, we consider the case where there are multiple antennas
at each of the three terminals, referring to it as the multi-input,
multi-output, multi-eavesdropper (MIMOME) channel.  In our model, the
channel matrices are fixed and known to all three terminals.  While
the eavesdropper's channel being known to both the sender and the
receiver in the problem formulation is a strong assumption, we remark
in advance that the solution provides ultimate limits on secure
transmission with multiple antennas, and thus serves as a starting
point for other formulations.  Further discussion of the modeling
assumptions is provided in the companion paper
\cite{khistiWornell:07} and the compound extension has been recently treated in~\cite{khisti:10}.

The problem of evaluating the secrecy capacity of channels with
multiple antennas has attracted increasing attention in recent years.  As
a starting point, for Gaussian models in which the channel matrices of
intended receiver and eavesdropper are square and diagonal, the
results in \cite{liangPoor07, liYatesTrappe:06a,
khistiTchamWornell:07, gopalaLaiElGamal:06Secrecy}, which consider
secure transmission over fading channels, can be applied.  In
particular, for this special case of independent parallel Gaussian
subchannels, it follows that using independent Gaussian wiretap
codebooks across the subchannels achieves capacity.

More generally, the MIMOME channel is a nondegraded broadcast channel
to which the Csisz{\'a}r-K{\"o}rner capacity expression
\cite{csiszarKorner:78} applies in principle.  However, computing the
capacity directly from \cite{csiszarKorner:78} appears difficult, as
observed in, e.g.,
\cite{ParadaBlahut:05,NegiGoel08,LiTrappeYates07,shaifeeUlukus:07}.

To the best of our knowledge, the first computable upper bound for the
secrecy capacity of the Gaussian multi-antenna wiretap channel appears
in \cite{khistiWornellEldar:07,khistiWornell:07}, which is used to
establish the secrecy capacity in the special (MISOME) case that the
intended receiver has a single antenna.  This approach involves
revealing the output of the eavesdropper's channel to the legitimate
receiver to create a fictitious degraded broadcast channel, and
results in a minimax expression for the upper bound, analogous to the
technique of Sato \cite{sato:78} used to upper bound the sum-capacity
of the multi-antenna broadcast channel; see, e.g., \cite{yuwei:06}.

In \cite{khistiWornellEldar:07,khistiWornell:07}, this minimax upper
bound is used to obtain a closed-form expression for the secrecy
capacity in the MISOME case.  In addition, a number of insights are
developed into the behavior of the secrecy capacity.  In the high
signal-to-noise ratio (SNR) regime, the simple masked beamforming
scheme developed in \cite{NegiGoel08} is shown to be near optimal.
Also, the scaling behavior of the secrecy capacity in the limit of
many antennas is studied.  

We note that this upper bounding approach has been independently
conceived by Ulukus et al.\ \cite{UlukusCorresp} and further applied
to the case of two transmit antennas, two receive antennas, and a
single eavesdropper antenna \cite{shafieeLiuUlukus:07}.  Subsequently,
this minimax upper bound was shown to be tight for the MIMOME case in
\cite{khistiWornell:07a} and, independently,
\cite{FrederiqueHassibi:07a} (see also \cite{FrederiqueHassibi:07}).
Both treatments start from the minimax upper bound of
\cite{khistiWornell:07} and work with the optimality conditions to
establish that the saddle value is achievable with the standard
Gaussian wiretap code construction \cite{csiszarKorner:78}.

In some of the most recent work, \cite{LiuShamai:07} provides an
alternative derivation of the MIMOME secrecy capacity using an
approach based on channel-enhancement techniques introduced in
\cite{weingartenSteinbergShamai:06}.  The two approaches shed
complementary insights into the problem.  The minimax upper bounding
approach in \cite{khistiWornell:07a,FrederiqueHassibi:07a} provides a
computable characterization for the capacity expression and identifies
a hidden convexity in optimizing the Csisz{\'a}r-K{\"o}rner expression
with Gaussian inputs, whereas the channel enhancement approach does
not.  On the other hand the latter approach establishes the capacity
given any covariance constraint on the input distribution, not just
the sum-power constraint to which the minimax upper bounding approach
has been limited.  

Finally, the diversity-multiplexing tradeoff of the
multi-antenna wiretap channel has been recently studied in
\cite{yukselErkip:08}.

An outline of the paper is as follows.  Section~\ref{sec:notation}
summarizes some notational conventions for the paper.
Section~\ref{sec:ChannelModel} describes the basic channel and system
model, as well as a canonical decomposition of the channel in terms of
its generalized singular values, which is used in some of the
asymptotic analysis.  Section~\ref{sec:Results} summarizes the main
results of the paper, and
Sections~\ref{sec:Capacity}--\ref{sec:Scaling} provide the
corresponding analysis.  In particular, Section~\ref{sec:Capacity}
develops the minimax characterization of the secrecy capacity,
Section~\ref{sec:HighSNR} develops the high SNR analysis in terms of
the generalized singular values, and Section~\ref{sec:Scaling}
develops the conditions under which the secrecy capacity is zero in
the limit of many antennas.  Finally, Section~\ref{sec:conclusions}
contains some concluding remarks.

\section{Notation}
\label{sec:notation}

In terms of fonts, bold upper and lower case characters are used for
matrices and vectors, respectively.  Random variables are
distinguished from their realizations by the use of san-serif fonts
for the former and regular serifed fonts for the latter.  Sets are
denoted using caligraphic fonts.  We generally reserve the symbols
$I(\cdot)$ for mutual information, and $h(\cdot)$ for differential
entropy, and all logarithms are base-2 unless otherwise indicated.  In
addition, $\CN(\bzero,\bK)$ denotes a circularly-symmetrix
complex-valued Gaussian random vector with covariance matrix $\bK$.

The set of all $n$-dimensional complex-valued vectors is denoted by
$\compls^n$, and the set of $m\times n$-dimensional matrices is
denoted using $\compls^{m\times n}$.  In addition, $\bI$ denotes the
identity matrix and $\bzero$ denotes the zero matrix.  When the
dimensions of these matrices is not clear from context, we will
explicily indicate their size via subscripts; e.g., $\bzero_{n \times
m}$ denotes an $n\times m$ zero matrix, $\bzero_n$ denotes a vector of
zeros of length $n$, and $\bI_n$ denotes an $n\times n$ identity
matrix.  We further use the notation $[\cdot]_{i:j}$ for $j\ge i$ to
denote the subvector of its vector argument corresponding to indices
$i,i+1,\dots,j$.  Likewise, $[\cdot]_{i:j,k:l}$ denotes the submatrix
formed from rows $i$ through $j$ and columns $k$ through $l$ of its
matrix argument.

Matrix transposition is denoted using the superscript ${\ }^\T$, the
Hermitian (i.e., conjugate) transpose of a matrix is denoted using the
superscript ${\ }^\dagger$, the Moore-Penrose pseudo-inverse is
denoted by ${\ }^\ddag$, and the projection matrix onto the null space
is denoted by ${\ }^\sharp$.  In addition, $\Null(\cdot)$,
$\rank(\cdot)$, and $\sigma_{\max}(\cdot)$ denote the null space,
rank, and largest singular value, respectively, of their matrix
arguments.  Moreover, we say a matrix has full column-rank if its rank
is equal to the number of columns, and the notation $\bA\succ\bzero$
means that $\bA$ is positive definite, with $\bA\succeq\bzero$ likewise
denoting positive semidefiniteness.

In other notation, $\dim(\cdot)$ denotes the dimension of its subspace
argument, $\spn(\cdot)$ denotes the subspace spanned by the collection
of vectors that are its argument, ${\ }^\perp$ denotes the orthogonal
complement of a subspace.  Moreover, $\|\cdot\|$ denotes the usual
Euclidean norm of a vector argument, $\tr(\cdot)$ and $\det(\cdot)$
denote the trace and determinant of a matrix, respectively, and
$\diag(\cdot)$ denotes a diagonal matrix whose diagonal elements are
given by its argument.

Finally, we use $\aseq$ and $\asconv$ to denote almost-sure equality
and convergence, respectively, and additionally use standard order
notation.  Specifically, $O(\eps)$ and $o(\eps)$ denote terms such
that $O(\eps)/\eps<\infty$ and $o(\eps)/\eps\rightarrow0$,
respectively, in the associated limit, so that, e.g., $o(1)$
represents a vanishing term.

\section{Channel and System Model}
\label{sec:ChannelModel} 

Using $\Nt$, $\Nr$, and $\Ne$ to denote the number of antennas at the
sender, intended receiver, and eavesdropper, respectively, the
received signals at the intended receiver and eavesdropper in the
channel model of interest are, respectively,
\begin{equation}
\begin{aligned}
\rvbyr(t) &= \bHr\rvbx(t) + \rvbzr(t)\\
\rvbye(t) &= \bHe\rvbx(t) + \rvbze(t)
\end{aligned}\ ,
\qquad t=1,2,\dots,n, 
\label{eq:channel}
\end{equation}
where $\rvbx(t)$ is the transmitted signal, where $\bHr \in
\compls^{\Nr\times \Nt}$ and $\bHe \in \compls^{\Ne\times \Nt}$ are
complex channel gain matrices, and where $\rvbzr(t)$ and $\rvbze(t)$
are each independent and identically distributed (i.i.d.) noises whose
samples are $\CN(\bzero,\bI)$ random variables.  The channel matrices
are constant (over the transmission interval) and known to all the
three terminals.  Moreover, the channel input satisfies the power
constraint
\begin{equation*} 
E\left[\frac{1}{n}\sum_{t=1}^n \|\rvbx(t)\|^2\right] \le P.
\end{equation*}

A rate $R$ is achievable if there exists a sequence of length $n$
codes such that both the error probability at the intended receiver
and $I(\rvw;\rvbye^n)/n$ approach zero as $n\rightarrow \infty$.  The
secrecy capacity is the supremum of all achievable rates.

\subsection{Channel Decomposition}

For some of our analysis, it will be convenient to exploit the
generalized singular value decomposition (GSVD)
\cite{paigeSaunders:81, vanloan76} of the channel \eqref{eq:channel}.
To develop this decomposition, we first define the subspaces
\begin{subequations}
\begin{align}
\cSr &= \Null(\bHr)^\perp \cap \Null(\bHe) \label{eq:cSr-def}\\ 
\cSre &=\Null(\bHr)^\perp \cap \Null(\bHe)^\perp \label{eq:cSre-def}\\
\cSe &= \Null(\bHr) \cap \Null(\bHe)^\perp \label{eq:cSe-def}\\
\cSn &= \Null(\bHr) \cap \Null(\bHe), \label{eq:cSn-def}
\end{align}
\label{eq:cS-def}%
\end{subequations}
corresponding to classes of inputs that have nonzero gain to,
respectively, the intended receiver only, both intended receiver and
eavesdropper, the eavesdropper only, and neither.  Letting
\begin{equation}
k \defeq \rank(\bHt)  
\label{eq:k-def}
\end{equation}
with
\begin{equation}
\bHt = \begin{bmatrix} \bHr \\ \bHe\end{bmatrix},
\label{eq:bHt-def}
\end{equation}
it follows that $\dim(\cSn)=\Nt-k$.  Moreover, we use the notation
\begin{equation} 
p \defeq \dim(\cSr) \qquad\text{and}\qquad s \defeq \dim(\cSre),
\label{eq:p-s-def}
\end{equation}
from which it follows that $\dim(\cSe)=k-p-s$.

Using this notation, our channel decomposition is as follows.
\begin{defn}
\label{defn:gsvd}
The GSVD of $(\bHr,\bHe)$ takes the form
\begin{subequations}
\begin{align}
\bHr &= 
\bPsir 
\bSir \begin{bmatrix} \bOm^{-1} & \bzero_{k\times(\Nt-k)} \end{bmatrix}
\bPsit^\dagger \label{eq:HrGsvd} \\
\bHe &= 
\bPsie 
\bSie \begin{bmatrix} \bOm^{-1} & \bzero_{k\times(\Nt-k)} \end{bmatrix}
\bPsit^\dagger, \label{eq:HeGsvd}
\end{align}%
\label{eq:gsvd-def}%
\end{subequations}
where $\bPsir \in \compls^{\Nr\times \Nr}$, $\bPsie \in \compls^{\Ne
  \times \Ne}$ and $\bPsit\in \compls^{\Nt\times \Nt}$ are unitary, 
where $\bOm \in \compls^{k\times
  k}$
is lower triangular and  nonsingular, and where
\begin{subequations}
\begin{align}
\bSir = {\kbordermatrix{& k-p-s & s & p\\
                         \Nr-p-s & \bzero & \bzero & \bzero \\
                         s & \bzero & \bDr & \bzero \\
                         p  & \bzero & \bzero & \bI}}
\label{eq:Sir-defn} \\
\bSie = {\kbordermatrix{& k-p-s & s & p\\
                         k-p-s & \bI & \bzero & \bzero \\
                         s  & \bzero & \bDev & \bzero \\
                         \Ne+p-k & \bzero & \bzero &\bzero}},
\label{eq:Sie-defn}
\end{align}%
\label{eq:Sire-defn}%
\end{subequations}
are diagonal with
\begin{equation}
\bDr = \diag(r_1,\ldots,r_s), \quad \bDev = \diag(e_1,\ldots,e_s),
\label{eq:Dre-defns}
\end{equation}
the diagonal entries of which are real and strictly positive.  The
associated generalized singular values are
\begin{equation}
\sigma_i \defeq \frac{r_i}{e_i}, \quad i=1,2,\dots, s.
\label{eq:GSV-defn}
\end{equation}
\end{defn}
For convenience, we choose the (otherwise arbitrary) indexing so that
$\sigma_1\le\sigma_2\le\dots\le\sigma_s$.

\section{Summary of Main Results}
\label{sec:Results}

In this section we summarize the main results in this paper.  The
analysis is provided in
Sections~\ref{sec:Capacity}--\ref{sec:Scaling}.

\subsection{MIMOME Secrecy Capacity}
\label{subsec:Capacity}

A characterization of the secrecy capacity of the MIMOME channel is as
follows.
\begin{thm}
The secrecy capacity of the MIMOME wiretap channel \eqref{eq:channel} is
\begin{equation}
C = \min_{\bK_\bPh \in \cK_\bPh}\max_{\bK_P\in \cK_P}
\Rp(\bK_P,\bK_\bPh), 
\label{eq:capacity}
\end{equation}
where 
\begin{equation} 
\Rp(\bK_P,\bK_\bPh) = I(\rvbx;\rvbyr|\rvbye),
\label{eq:Rplus-def}
\end{equation}
with $\rvbx \sim \CN(\bzero,\bK_P)$ and
\begin{equation}
\cK_P \defeq \bigl\{ \bK_P \,:\, \bK_P \succeq \bzero,\quad
\tr(\bK_P) \le P \bigr\},
\label{eq:KP-def}
\end{equation}
and where 
\begin{equation} 
\rvbz \defeq \begin{bmatrix} \rvbzr \\ \rvbze \end{bmatrix} 
\sim \CN(\bzero,\bK_\bPh), 
\label{eq:z-def}
\end{equation}
with\footnote{The constraint $\bK_\bPh\succeq\bzero$ is equivalently
  expressed as the requirement that $\sigma_{\max}(\bPh)\le1$, as we
  will exploit.}
\begin {align}
\cK_\bPh 
&\defeq \left\{\bK_\bPh \,:\, \bK_\bPh 
= \begin{bmatrix} \bI_\Nr & \bPh \\ \bPh^\dagger & \bI_\Ne \end{bmatrix},
\ \bK_\bPh \succeq \bzero \right\}. \label{eq:Kph-def}
\end{align}
Furthermore, the minimax problem of \eqref{eq:capacity} is
convex-concave with saddle point solution $(\obK_P,\obK_\bPh)$, 
via which the secrecy capacity can be expressed in the form
\begin{equation} 
C 
= \Rm(\obK_P) 
\defeq \log\frac{\det(\bI + \bHr\obK_P\bHr^\dagger)}{\det(\bI +
  \bHe\obK_P\bHe^\dagger)}. 
\label{eq:Rm-def}
\end{equation}
Finally, $C=0$ if and only if 
\begin{equation} 
\bHr=\obTh\bHe,
\label{eq:zero-cap-cond}
\end{equation}
where
\begin{equation} 
\obTh \defeq \bTh(\obK_P), \qquad \bTh(\bK_P) \defeq
\bTh(\bK_P,\obK_\bPh), 
\label{eq:obTh-def}
\end{equation}
with 
\begin{equation}
\bTh(\bK_P,\bK_\bPh) \defeq (\bHr\bK_P\bHe^\dagger + \bPh)(\bI +
\bHe\bK_P\bHe^\dagger)^{-1}
\label{eq:bTh-def}
\end{equation}
denoting the coefficient in the linear minimum mean-square error
(MMSE) estimate of $\rvbyr$ from $\rvbye$, 
\label{thm1}
\end{thm}

\iffalse
As an initial remark, note that \eqref{eq:capacity} does not express
the capacity in closed-form, the capacity can be readily computed
numerically in this form.
\fi

Several remarks are worthwhile.  First, our result can be related to
the Csisz\'{a}r-K\"{o}rner characterization of the secrecy capacity
for a nondegraded discrete memoryless broadcast channel
$p_{\rvyr,\rvye|\rvx}$ in the form \cite{csiszarKorner:78}
\begin{align}
C = \max_{p_{\rvu},p_{\rvx|\rvu}} I(\rvu;\rvyr)-I(\rvu;\rvye),
\label{eq:CK}
\end{align}
where $\rvu$ is an auxiliary random variable (over some alphabet with
bounded cardinality) that satisfies the Markov constraint $\rvu
\leftrightarrow \rvx \leftrightarrow (\rvyr,\rvye)$.  As
\cite{csiszarKorner:78} remarks, the secrecy capacity \eqref{eq:CK}
can be extended to incorporate continuous-valued inputs of the type of
interest in the present paper.  With such an extension,
Theorem~\ref{thm1}, and in particular \eqref{eq:Rm-def}, can be
interpreted as (indirectly) establishing a suitable Gaussian wiretap
code for achieving capacity.\footnote{Each candidate
$(\rvu,\rvx)$ in \eqref{eq:CK} corresponds to a particular coding
scheme based on binning, which we generically refer as a ``wiretap
code,'' which achieves rate $I(\rvu;\rvyr)-I(\rvu;\rvye)$. }
Specifically, via the chain rule,
\begin{equation*} 
I(\rvbx;\rvbyr|\rvbye) = \bigl[ I(\rvbx;\rvbyr) - I(\rvbx;\rvbye)
  \bigr] + I(\rvbx;\rvbye|\rvbyr)
\end{equation*}
where the last term on the right-hand side is zero when $\bPh=\obPh$,
and thus we have the following immediate corollary.
\begin{corol}
The secrecy capacity of the MIMOME wiretap channel is achieved by a
wiretap coding scheme in which $\rvbu\sim \CN(\bzero,\bK_P)$ with
$\bK_P=\obK_P$, and $\rvbx=\rvbu$.
\end{corol}
\iffalse
In particular, via a straightforward calculation we note that using
such choices this scheme achieves rate
\begin{align} 
\Rm(\obK_P) 
&\defeq I(\rvbu;\rvbyr)-I(\rvbu;\rvbye) \notag\\
&= \log\frac{\det(\bI +\bHr\obK_P\bHr^\dagger)}{\det(\bI +
  \bHe\obK_P\bHe^\dagger)}, \label{eq:Rm-def}
\end{align}
which coincides with \eqref{eq:alternateCap}.
\fi

From this perspective, our result can also be interpreted as a convex
reformulation of the nonconvex optimization \eqref{eq:CK}.  Indeed,
even after knowing that both an optimizing $\rvbu$ is Gaussian and
$\rvbx=\rvbu$ is sufficient---which itself is nontrivial---determining
the optimal covariance via
\begin{equation}
\label{eq:achievRate}
\obK_P \in \argmax_{\bK_P \in \cK_P}\log\frac{\det(\bI + \bHr
  \bK_P \bHr^\dagger)}{\det(\bI + \bHe \bK_P \bHe^\dagger)}
\end{equation}
with $\cK_P$ as defined in \eqref{eq:KP-def}, is a nonconvex
problem.\footnote{Note that in the high-SNR regime,
\eqref{eq:achievRate} reduces to
\begin{equation*} 
\max_{\bK \in \cK_{\infty}}
\log \frac{\det(\bHr\bK \bHr^\dagger)}{\det(\bHe\bK\bHe^\dagger)},
\end{equation*}
which is the well-studied multiple-discriminant function in
multivariate statistics; see, e.g., \cite{Wilks:62}.}  And even if one
verifies that $\obK_P$ satisfies the Karush-Kuhn-Tucker (KKT)
conditions associated with \eqref{eq:achievRate}, these necessary
conditions only establish local optimality, i.e., that $\obK_P$ is a
stationary point of the associated objective function.  By contrast,
\eqref{eq:capacity} establishes that the (global) solution to
\eqref{eq:achievRate} is obtained as the solution to a convex problem,
as well as establishing the optimality of a Gaussian input
distribution.

Second, additional insights are obtained from the
structure of the saddle point solution $(\obK_P,\obK_\bPh)$.  In
particular, using $\obPh$ to denote the optimal cross-covariance, i.e.,
[cf.~\eqref{eq:Kph-def}]
\begin{equation} 
\obK_\bPh \defeq \bK_\obPh 
= \begin{bmatrix} \bI_\Nr & \obPh\\\obPh^\dagger & \bI_\Ne \end{bmatrix},
\label{eq:oPh-def} 
\end{equation}
we establish in the course of our development of Theorem~\ref{thm1} the
following key property.
\begin{property}
\label{prop:degradedness}
The saddle point solution $(\obK_P,\obK_\bPh)$ to the MIMOME wiretap
channel capacity \eqref{eq:capacity} satisfies
\begin{equation}
\obPh^\dagger \bHr \obS = \bHe\obS,\text{ $\forall\!$ full column-rank
  $\obS$ s.t.\ $\obS\obS^\dagger=\obK_P$,} 
\label{eq:degradedness}
\end{equation}
provided $\bHr\neq\obTh\bHe$ (i.e., provided $C\neq0$).
\end{property}
It follows from \eqref{eq:degradedness} that the effective channel to
the eavesdropper is a degraded version of that to the intended
receiver.  Indeed, the intended receiver can simulate the eavesdropper
channel by adding noise.  Specifically, it generates
\begin{equation*}
\rvbye' = \obPh^\dagger \rvbyr + \rvbw,
\end{equation*}
where the added noise $\rvbw\sim \CN(\bzero,\bI-\obPh^\dagger\obPh)$
is independent of $\rvbyr$, so, using \eqref{eq:channel},
\eqref{eq:degradedness}, and the notation $\rvbx=\obS\rvbx'$ with
$\rvbx'\sim\CN(\bzero,\bI)$, we have
\begin{align*}
\rvbye' = \obPh^\dagger \bHr \obS \rvbx' + \obPh^\dagger \rvbzr +
\rvbw = \bHe\obS\rvbx' + \rvbze' = \bHe\rvbx + \rvbze',
\end{align*}
where $\rvbze'\sim\CN(\bzero,\bI)$.  In essence, the optimal signal
design for transmission is such that no information is transmitted
along any direction where the eavesdropper observes a stronger signal
than the legitimate receiver.  A key consequence is that a genie-aided
system in which $\rvbye$ is provided to the receiver, which would
otherwise provide only an upper bound on capacity in general, does not
increase the capacity of the channel in this case, a feature that is
ultimately central to our analysis.

Finally, the condition \eqref{eq:zero-cap-cond} corresponding to when
the secrecy capacity is zero has a natural physical interpretation.
In particular, under this condition, the effective channel to the
intended receiver is a degraded version of that to the eavesdropper.
Indeed, the eavesdropper can simulate the intended receiver by adding
noise.   Specifically, it generates
\begin{equation*} 
\rvbyr' = \obTh \rvbye + \rvbw,
\end{equation*}
where the added noise $\rvbw\sim \CN(\bzero,\bI-\obPh\obPh^\dagger)$
is independent of $\rvbyr$, so, using \eqref{eq:channel} we have
\begin{align*}
\rvbyr' = \obTh \bHe \rvbx + \obTh \rvbzr +
\rvbw = \bHr\rvbx + \rvbzr',
\end{align*}
where $\rvbzr'\sim\CN(\bzero,\bI)$ since 
\begin{equation} 
\obTh=\obPh \quad\text{if}\quad \bHr =\obTh\bHe,
\label{eq:zero-cap-obTh}
\end{equation}
which follows from \eqref{eq:obTh-def} with \eqref{eq:bTh-def}.

\subsection{Secrecy Capacity in the High-SNR Regime}
\label{subsec:HighSNR}

In the high-SNR limit (i.e., $P\rightarrow\infty$), the secrecy
capacity \eqref{eq:capacity} is naturally described in terms of the
GSVD of the channel \eqref{eq:channel} as defined in
\eqref{eq:gsvd-def}.  The GSVD simultaneously diagonalizes the $\bHr$
and $\bHe$, yielding an equivalent parallel channel model for the
problem.  As such, a capacity-approaching scheme in the high-SNR
regime involves using for transmission (with a wiretap code) only
those subchannels for which the gain to the intended receiver is
larger, and the following convenient expression for the capacity
\eqref{eq:capacity} results.
\begin{thm}
Let $\sigma_1 \le \sigma_2 \le \ldots \le \sigma_{s}$ be the
generalized singular values of $(\bHr,\bHe)$.  Then as
$P\rightarrow\infty$, the secrecy capacity of the MIMOME wiretap
channel \eqref{eq:channel} takes the asymptotic form
\begin{equation}
C(P) = C_0(P) + \sum_{j\,:\,\sigma_j \ge 1} \log \sigma_j^2 - o(1),
\label{eq:highSNRCapGen}
\end{equation}
where
\begin{equation}
C_0(P) = \begin{cases} \ds\log\det\Bigl(\bI + \frac{P}{p} \bHr \bHe^\sharp
\bHr^\dagger\Bigr), & \rank(\bHe)<\Nt, \\
0, & \rank(\bHe)=\Nt, \end{cases}
\label{eq:cop-def}
\end{equation}
with $p$ and $s$ as given in \eqref{eq:p-s-def}, and with $\bHe^\sharp$
denoting the projection matrix onto $\Null(\bHe)$.
\label{thm:HighSNRCap}
\end{thm}

Note that a simple and intuitive transmission scheme for the MIMOME
channel would involve simultaneously and isotropically transmitting
information in $\Null(\bHr)^\perp$, where there is gain to the
intended receiver, and (synthetic) noise in $\Null(\bHr)$, which does
not affect the intended receiver but does reduce the quality of the
eavesdroppers received signal.\footnote{Note that the scheme is
semi-blind: the transmitter does not need to know $\bHe$ to construct
the required subspaces, but does need to know $\bHe$ in order to
choose the communication rate.}  This ``masked'' multi-input,
multi-output (MIMO) transmission scheme is the natural generalization
of the masked beamforming proposed in \cite{NegiGoel08} for the MISOME
wiretap channel.  For the MISOME channel, such an approach is near
optimal, as shown in \cite{khistiWornell:07}.  However, we now show
that such a masked multi-input multi-output (MIMO) scheme can be quite
far from optimal on the MIMOME channel.

For convenience, we restrict our attention to the case in
which $\Nr\le\Nt\le\Ne$ and $\bHr$ and $\bHe$ are full rank---i.e.,
$\rank(\bHr)=\Nr$ and $\rank(\bHe)=\Nt$---and thus $k=\Nt$, $p=0$, and
$s=\Nr$ in the GSVD.

The masked MIMO scheme is naturally viewed as a wiretap coding scheme
in which a particular (rather than optimal) choice for $(\rvbx,\rvbu)$
is imposed in \eqref{eq:CK}.  In particular, first we choose $\rvbu$
to correspond to (information-bearing) codewords in a randomly
generated codebook, i.e.,
\begin{subequations} 
\begin{equation} 
\rvbu=(\rvb_1,\dots,\rvb_\Nr),
\label{eq:rvbu-sn}
\end{equation}
where the elements are generated in an i.i.d.\ manner according to
$\CN(0,\Pt)$ with 
\begin{equation} 
\Pt \defeq \frac{P}{\Nt}.
\label{eq:Pt-def}
\end{equation}
Additionally, we let $\rvb_{\Nr+1},\ldots, \rvb_\Nt$ be randomly
generated (synthetic) noise, i.e., independent $\CN(0,\Pt)$ random
variables.

Next, we choose the transmission $\rvbx$ according to
\begin{equation}
\rvbx = \sum_{j=1}^{\Nt} \rvb_j \bv_j,
\label{eq:rvbx-sn}
\end{equation}%
\label{eq:SyntheticNoiseParams}%
\end{subequations}
where the vectors $\bv_1,\dots,\bv_\Nt$ are chosen as follows.  Let
\begin{equation}
\bHr = \bU \bHrsv \bVr^\dagger
\label{eq:bHr-svd}
\end{equation}
be the compact singular value decomposition (SVD) of $\bHr$.  Since
$\rank(\bHr)= \Nr$, this means that $\bU$ is $\Nr\times\Nr$ and
unitary, $\bHrsv$ is $\Nr\times\Nr$ and diagonal with positive diagonal
elements, and $\bVr$ is $\Nt\times\Nr$ with orthogonal columns.  Then
we choose $\bv_1,\dots,\bv_\Nr$ in \eqref{eq:rvbx-sn} as the columns
of $\bVr$, i.e., 
\begin{equation*} 
\bVr = \begin{bmatrix} \bv_1 & \bv_2 & \cdots & \bv_\Nr \end{bmatrix},
\end{equation*}
and (freely) choose 
\begin{equation} 
\bVn \defeq 
\begin{bmatrix} \bv_{\Nr+1} & \cdots &\bv_{\Nt} \end{bmatrix},
\label{eq:bVn-def}
\end{equation}
a basis for the null space of $\bHr$, so that $\begin{bmatrix}
\bVr & \bVn \end{bmatrix}$ is unitary.

As we will establish, substituting these parameters in the argument of
\eqref{eq:CK} yields the achievable rate
\begin{equation} 
R_\mrm{SN}(P) = \log\det\left[\bigl(\Pt\bI + \bHrsv^{-2}\bigr)
\bigl(\bHr(\bI + \Pt\bHe^\dagger\bHe)^{-1}\bHr^\dagger\bigr)\right],
\label{eq:Rsn}
\end{equation}
which in the high-SNR regime reduces to
\begin{equation}
\lim_{P\rightarrow\infty}R_\mrm{SN}(P) =
\log\det\bigl(\bHr(\bHe^\dagger\bHe)^{-1}\bHr^\dagger\bigr) 
= \sum_{j=1}^{\Nr}\log\sigma_j^2,
\label{eq:Rsn_highSNR}
\end{equation}
where the second equality comes from expanding $\bHr$ and $\bHe$ via
\eqref{eq:gsvd-def}, with $\sigma_1,\sigma_2,\ldots$ denoting the
generalized singular values \eqref{eq:GSV-defn}.  Comparing
\eqref{eq:Rsn_highSNR} and \eqref{eq:highSNRCapGen}, we see that the
asymptotic gap to capacity is
\begin{equation*} 
\lim_{P\rightarrow\infty} \bigl[ C(P)-R_\mrm{SN}(P) \bigr] 
= \sum_{j\,:\,\sigma_j<1} \log\frac{1}{\sigma_j^2},
\end{equation*}
which, evidently, can be arbitrarily large when there are small
singular values.

In concluding this section, we emphasize that only in the high-SNR
regime do the generalized singular values of $(\bHr,\bHe)$ completely
characterize the capacity-achieving and masked MIMO coding schemes.

\subsection{MIMOME Channel Scaling Laws}
\label{subsec:Scaling} 

By using sufficiently many antennas, the eavesdropper can drive to
secrecy capacity to zero.  In such a regime, the eavesdropper would be
able to decode a nonvanishing fraction of any sent message---even when
the sender and receiver fully exploit knowledge of $\rvbHe$.  In
general, this threshold depends on the numbers of antennas at the
transmitter and intended receiver, as well as on the particular
channels to intended receiver and eavesdropper.  One characterization
of this threshold is given by \eqref{eq:zero-cap-cond} in
Theorem~\ref{thm1}.  An equivalent characterization that is more
useful in the development of scaling laws, is as follows.
\begin{claim}
The secrecy capacity of the MIMOME channel is zero if and only if
\begin{equation}
\sigma_{\max}(\bHr,\bHe) \defeq
\sup_{\bv\in\compls^{\Nt}}\frac{\|\bHr\bv\|}{\|\bHe\bv\|} \le 1.  
\label{eq:largeEigValue}
\end{equation}
where $\sigma_{\max}(\bHr,\bHe)$ denotes the channel's largest
generalized singular value.
\label{lem:zeroCapCond}
\end{claim}

When the coefficients of the channels are drawn at random, and the
numbers of antennas are large, the threshold becomes independent of
the channel realization.  The following result characterizes this
scaling behavior.
\begin{corol}
Suppose that $\rvbHr$ and $\rvbHe$ have i.i.d.\ $\CN(0,1)$ entries
that are fixed for the entire period of transmission, and known to all
the terminals.  Then when $\Nr,\Ne,\Nt \rightarrow \infty$ such that
$\g\defeq \Nr/\Ne$ and $\bt\defeq\Nt/\Ne$ are fixed constants, the
secrecy capacity satisfies $C(\rvbHr,\rvbHe)\asconv0$ if and only if
\begin{equation}
0 \le \bt \le \frac{1}{2} \quad\text{and}\quad \g \le (1-\sqrt{2\bt})^2.
\label{eq:AsympZeroCondn}
\end{equation}
\label{corol:AsympZeroCondn}
\end{corol}

\begin{figure}[tbp]
\includegraphics[width=3.5in]{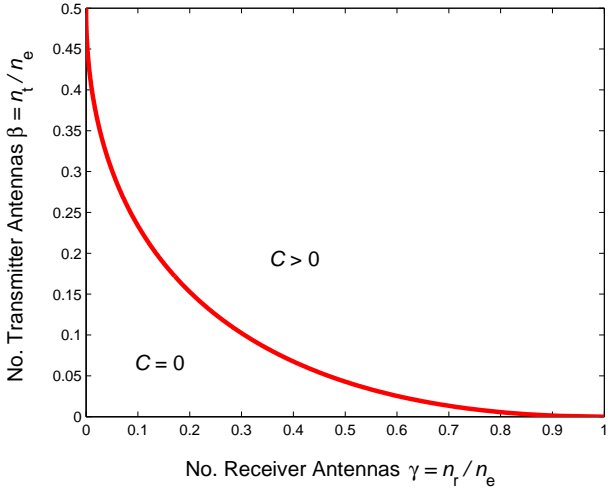}
\caption{The efficient frontier of secure communication region as a
  function of the number of antennas at the transmitter and intended
  receiver (relative to the number at the eavesdropper), in the
  limit of many antennas.  The capacity is zero for any point below
  the curve, i.e., whenever the eavesdropper has sufficiently many
  antennas. \label{fig:zeroPlot1}}
\end{figure}

Fig.~\ref{fig:zeroPlot1} depicts the zero-capacity region
\eqref{eq:AsympZeroCondn}.  In this plot, the solid curve describes
the relative number of antennas an eavesdropper needs to prevent
secure communication, as a function of the antenna resources available
at the transmitter and intended receiver.  The related scaling law
developed for the MISOME case \cite{khistiWornell:07} corresponds to
the vertical intercept of this plot: $C\asconv0$ when $\bt \le 1/2$,
i.e., when the eavesdropper has at least twice the number of antennas
as the sender.  Note, too, that the single transmit antenna (SIMOME)
case corresponds to the horizontal intercept; in this case we see that
$C\asconv0$ when $\g \le 1$, i.e., when the eavesdropper has more
antennas than the intended receiver.

We can further use such scaling analysis to determine the best asymptotic
allocation of a (large) fixed number of antennas $T$ between
transmitter and intended receiver in the presence of an an
eavesdropper.    In particular, the optimum allocation is
\begin{equation} 
(\bt_*,\g_*) = \!\!\!\!\argmin_{\substack{
\bigl\{(\bt,\g)\,:\ 0\le\bt\le1/2,\\ \qquad\quad 0\le\g\le(1-\sqrt{2\bt})^2\bigr\}}}
\!\!\!\!(\bt+\g) = \left(\frac{2}{9},\frac{1}{9}\right),
\label{eq:myOpt}%
\end{equation}
as is easily verified.  Thus, the allocation that best thwarts the
eavesdropper is $\Nr/\Nt=1/2$, which requires the eavesdropper to use
$3T$ antennas to prevent secure communication.

\begin{figure}[tbp]
\includegraphics[width=3.5in]{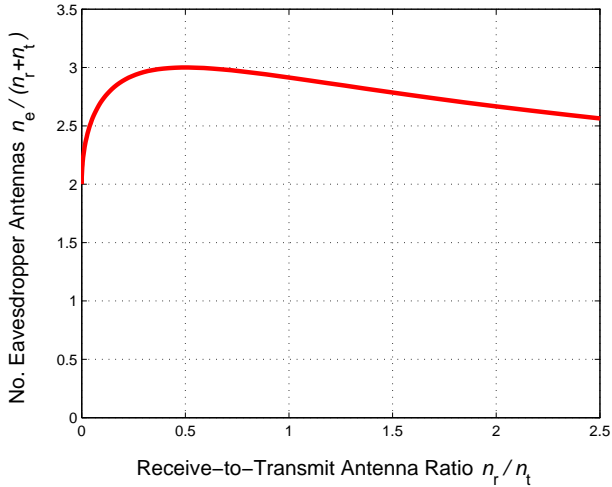}
\caption{The minimum (relative) number of eavesdropper antennas
  required to drive the secrecy capacity to zero, as a function of the
  antenna allocation between transmitter and intended receiver, in the
  limit of many antennas. }
\label{fig:zeroPlot2}
\end{figure}

It is worth remarking that the objective function in \eqref{eq:myOpt}
is rather insensitive to deviations from the optimal antenna
allocation, as Fig.~\ref{fig:zeroPlot2} demonstrates.  If fact, even
if we were to allocate equal numbers of antennas to the sender and the
receiver, the eavesdropper would still need $(3/2+\sqrt{2})T \approx
2.9142\,T$ antennas to drive the secrecy capacity to zero.

\section{MIMOME Secrecy Capacity Analysis}
\label{sec:Capacity}

In this section we prove Theorem~\ref{thm1}.  Our proof involves two
main parts.  We first recognize the right-hand side of
\eqref{eq:capacity} as an upper bound on the secrecy capacity, then
exploit properties of the saddle point solution to establish
\begin{equation}
\Rp(\obK_P,\obK_\bPh)=\Rm(\obK_P),
\label{eq:SaddlePointProperty}
\end{equation}
where $\Rm(\obK_P)$ is the lower bound (achievable rate) given in
\eqref{eq:Rm-def}.

We begin by stating our upper bound, which is a trivial generalization
of that established in \cite{khistiWornell:07}.
\begin{lemma}[\cite{khistiWornell:07}]
\label{lem:UpperBound}
An upper bound on the secrecy capacity of the MIMOME channel
\eqref{eq:channel} is given by
\begin{equation}
C(P) \le \Rub = \min_{\bK_\bPh \in \cK_\bPh}\max_{\bK_P\in \cK_P}\Rp(\bK_P,\bK_\bPh),
\label{eq:capacityUB}
\end{equation}
where 
\begin{equation}
\Rp(\bK_P,\bK_\bPh)\defeq I(\rvbx;\rvbyr|\rvbye),
\label{eq:condMutInf}
\end{equation}
with $\rvbx \sim \CN(\bzero,\bK_P)$, and $\rvbz \sim
\CN(\bzero,\bK_\bPh)$, and the domain sets $\cK_P$ and $\cK_\bPh$
are defined via \eqref{eq:KP-def} and \eqref{eq:Kph-def} respectively.
\end{lemma}

\begin{figure}[tbp]
\small
\psfrag{Sdl}{\ Saddle Point: $(\obK_P,\obK_\bPh)$}
\psfrag{Kcond}{$\displaystyle \obK_\bPh \in \argmin_{\!\cK_\bPh}\!\Rp(\!\obK_P,\!\bK_\bPh\!)$}
\psfrag{Qcond}{\,$\displaystyle\obK_P \in \argmax_{ \cK_P} \Rp(\bK_P,\obK_\bPh)$}
\psfrag{Qcond2}{\ \,$\displaystyle\obK_P \in \argmax_{ \cK_P} h(\rvbyr-\obTh\rvbye)$}
\psfrag{FinalCondn}{\ $\obPh^\dagger \bHr \obS =  \bHe \obS \Rightarrow \Rp(\obK_P,\obK_\bPh)=\Rm(\obK_P)$}
\includegraphics[scale=.34]{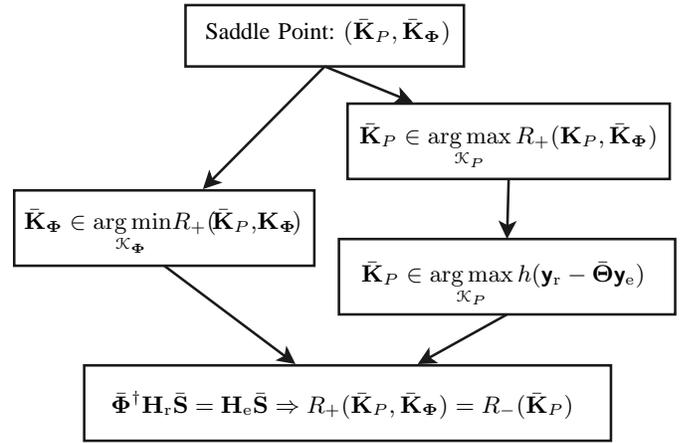}
\caption{Key steps in the proof of Theorem~\ref{thm1}.  First, the existence of
  a saddle point $(\obK_P,\obK_\bPh)$ is established, then the KKT
  conditions associated with the minimax expressions are used to
  simplify the saddle value to show that it matches the lower bound.
  \label{fig:proof}}
\end{figure}

It remains to establish that this upper bound expression
satisfies \eqref{eq:SaddlePointProperty}, which we do in the
remainder of this section.   We divide the proof into several steps,
as depicted in Fig.~\ref{fig:proof}.

Furthermore, we remark in advance that the analysis throughout is
slightly simpler when $\bK_\bPh\succ\bzero$.  Accordingly, in the
following sections we focus on this nonsingular case and defer analysis
for the singular case to appendices as it arises in our development.
The key to analysis of the singular case is replacing the observations
$\rvbyr$ with reduced but equivalent observations.  In particular, we
will make use of the following claim, a proof of which is provided in
Appendix~\ref{app:equiv-obs}.
\begin{claim}
\label{claim:noiseSing}
Let the singular value decomposition of $\bPh$ be expressed the
form
\begin{equation}
\bPh = \begin{bmatrix} \bU_1 & \bU_2\end{bmatrix} 
\begin{bmatrix}\bI &
                       \bzero\\\bzero & \bPhsv\end{bmatrix} 
\begin{bmatrix}\bV_1^\dagger \\\bV_2^\dagger\end{bmatrix},\quad
\sigma_{\max}(\bPhsv)< 1.
\label{eq:oPhSVD}
\end{equation}
Then if $p_\rvbx$ is such that $I(\rvbx;\rvbyr | \rvbye ) < \infty$,
we have
\begin{equation}
I(\rvbx;\rvbyr|\rvbye) = I(\rvbx;\rvbytr| \rvbye), \label{eq:noiseSing3a}
\end{equation}
where 
\begin{equation} 
\rvbytr \defeq \bU_2^\dagger\rvbyr = \bHtr \rvbx + \rvbztr
\label{eq:rvbytr-def}
\end{equation}
with 
\begin{equation} 
\bHtr \defeq \bU_2^\dagger\bHr \quad\text{and}\quad \rvbztr \defeq
\bU_2^\dagger\rvbzr \label{eq:bHtr-rvbztr-def}.
\end{equation}
Symmetrically, if $p_\rvbx$ is such that $I(\rvbx;\rvbye | \rvbyr ) < \infty$,
we have
\begin{equation} 
I(\rvbx;\rvbye|\rvbyr) = I(\rvbx;\rvbyte| \rvbyr), \label{eq:noiseSing3b}
\end{equation}
where 
\begin{equation} 
\rvbyte \defeq \bV_2^\dagger\rvbye = \bHte \rvbx + \rvbzte
\label{eq:rvbyte-def}
\end{equation}
with 
\begin{equation} 
\bHte \defeq \bV_2^\dagger\bHe \quad\text{and}\quad \rvbzte \defeq
\bV_2^\dagger\rvbze. \label{eq:bHte-rvbzte-def}
\end{equation}
Finally, for any $p_\rvbx$ we have that
$I(\rvbx;\rvbyr|\rvbye)=\infty$ if and only if
%if $p_\rvbx$ is such that $I(\rvbx;\rvbyr|\rvbye) =
%\infty$, then either \eqref{eq:noiseSing3a} holds or
\begin{equation}
\bT \bK_P \bT^\dagger \neq \bzero,
\label{eq:inf-I-cond}
\end{equation}
where $\bK_P$ is the covariance associated with $p_\rvbx$, and where
\begin{equation} 
\bT \defeq \bU_1^\dagger\bHr-\bV_1^\dagger\bHe.
\label{eq:bT-def}
\end{equation}
%Finally, for any $p_\rvbx$ such that \eqref{eq:inf-I-cond} holds,
%$I(\rvbx;\rvbyr|\rvbye) = \infty$.
\end{claim}

Note that when \eqref{eq:noiseSing3a} holds, the equivalent model
holds and
\begin{equation}
\tbPhr \defeq E[\rvbztr\rvbze^\dagger] = \bU_2^\dagger \bPh
\label{eq:equiv-noise-cov}
\end{equation}
is the equivalent noise cross-covariance.

\subsection{Existence of a Saddle Point Solution}

We first show that the minimax upper bound is a convex-concave problem
with a (finite) saddle point solution.
\begin{lemma}
\label{lem:saddlePointExistence}
The upper bound \eqref{eq:capacityUB} has a saddle point solution, i.e.,
there exists  $(\obK_P,\obK_\bPh) \in \cK_P\times\cK_\bPh$ such that
\begin{equation}
\Rp(\bK_P,\obK_\bPh) \le \Rp(\obK_P,\obK_\bPh) \le \Rp(\obK_P,\bK_\bPh)
\label{eq:saddlePointProp}
\end{equation}
holds for each $(\bK_P,\bK_\bPh) \in \cK_P\times\cK_\bPh$.   Moreover, 
the saddle value is finite, i.e.,
\begin{equation} 
\Rp(\obK_P,\obK_\bPh)<\infty.
\label{eq:finite-saddle}
\end{equation}
\end{lemma}

\begin{IEEEproof}
Since the constraint sets $\cK_P$ and $\cK_\bPh$ are convex and
compact, from a special case of Sion's minimax theorem \cite{sion} it
suffices to show that
\begin{gather}
\text{$\Rp(\bK_P,\cdot)$ is convex on $\cK_\bPh$ for each $\bK_P\in \cK_P$}
\tag{P1} \label{p:convex} \\
\text{$\Rp(\cdot,\bK_\bPh)$ is
concave on $\cK_P$ for each  $\bK_\bPh \in \cK_\bPh$}
\tag{P2} \label{p:concave}
\end{gather}

To first establish \eqref{p:convex}, we begin by writing
\begin{equation} 
I(\rvbx;\rvbyr|\rvbye)= I(\rvbx;\rvbyr,\rvbye)-I(\rvbx;\rvbye),
\label{eq:I-exp}
\end{equation}
and observe that the second term in \eqref{eq:I-exp} is fixed for each
$\bK_\bPh \in \cK_\bPh$.  Thus it suffices to show that with $\rvbx
\sim \CN(\bzero,\bK_P)$, the first term in \eqref{eq:I-exp} is convex in
$\bK_\bPh$.  This is established in, e.g., \cite[Lemma~II-3,
  p.~3076]{diggaviCover01}. 

We next establish \eqref{p:concave}.  With slight abuse of notation, we
define $\Rp(p_\rvbx, \bK_\bPh) = I(\rvbx;\rvbyr| \rvbye)$ with
$\rvbx\sim p_\rvbx$ and $\rvbz\sim\CN(\bzero,\bK_\bPh)$.  By contrast,
our original notation $\Rp(\bQ, \bK_\bPh)$ corresponds to the special
case of $\Rp(p_\rvbx, \bK_\bPh)$ in which $p_\rvbx=\CN(\bzero,\bQ)$.
Let $p_\rvbx^0 = \CN(\bzero,\bQ_0)$, $p_\rvbx^1 = \CN(\bzero,\bQ_1)$,
$p_\rvbx^\theta = \theta p_\rvbx^1 +(1-\theta)p_\rvbx^0$, and
$\bQ^\theta = (1-\theta) \bQ_0 +\theta \bQ_1$, for some $\theta \in
[0,1]$.  Then the required concavity follows from
\begin{align}
\Rp(\bQ^\theta,\bK_\bPh) &= \Rp(\CN(\bzero,\bQ^\theta),\bK_\bPh) \notag\\
&\ge \Rp(p^\theta_\rvbx,\bK_\bPh) \label{eq:GaussMax}\\
&\ge (1-\theta) \Rp(p_\rvbx^0, \bK_\bPh) + \theta
\Rp(p_\rvbx^1,\bK_\bPh) \label{eq:Concavity}\\
&= (1-\theta) \Rp(\bQ_0, \bK_\bPh) + \theta \Rp(\bQ_1,\bK_\bPh), \notag
\end{align}
where \eqref{eq:GaussMax} follows from the fact that a Gaussian
distribution maximizes $\Rp(p_\rvbx^\theta,\bK_\bPh)$ among
all distributions with a given covariance, which we discuss below,
and where \eqref{eq:Concavity} follows from
the fact that $I(\rvbx;\rvbyr|\rvbye)$ is concave in 
$p_\rvbx$ for each fixed $p_{\rvbyr,\rvbye|\rvbx}$; see,
e.g., \cite[Appendix I]{khistiTchamWornell:07}.

Verifying \eqref{eq:GaussMax} is straightforward when $\bK_\bPh$ is
nonsingular, i.e., $\|\bPh\|_2 < 1$.  Specifically, with 
\begin{align}
&\bLa(\bK_P,\bK_\bPh) \notag\\
&\ \defeq \bI + \bHr\bK_P\bHr^\dagger \notag\\
&\ \quad {} - (\bPh + \bHr \bK_P\bHe^\dagger) 
(\bI +\bHe\bK_P\bHe^\dagger)^{-1} (\bPh^\dagger + \bHe\bK_P\bHr^\dagger)
\label{eq:bLa-def}
\end{align}
denoting the error covariance associated with the linear MMSE estimate
$\bTh(\bK_P,\bK_\bPh)\rvbye$ of $\rvbyr$ from $\rvbye$, a simple
generalization of \cite[Lemma~2]{khistiWornell:07} yields
\begin{align}
I(\rvbx;\rvbyr | \rvbye) 
&= h(\rvbyr | \rvbye) - h(\rvbzr | \rvbze) \label{eq:mutInfExp} \\ 
&= h(\rvbyr | \rvbye) - \log\det\pi e(\bI-\bPh\bPh^\dagger) \notag\\ 
&\le \log\det \bLa(\bK_P,\bK_\bPh) - \log\det(\bI-\bPh\bPh^\dagger),
\label{eq:mutInfExp2}
\end{align}
where the last inequality is satisfied with equality if $p_\rvbx =
\CN(\bzero,\bK_P)$.  When $\bK_\bPh$ is singular, \eqref{eq:mutInfExp}
is not well-defined, so some straightforward modifications to the
approach are required; these we detail in Appendix~\ref{GaussOpt}.

Finally, to verify \eqref{eq:finite-saddle}, it suffices to note
that 
\begin{equation*} 
\Rp(\obK_P,\obK_\bPh) \le \Rp(\obK_P,\bI) \le I(\rvbx;\rvbye,\rvbyr) <
\infty
\end{equation*}
where the second inequality follows from the chain rule
$I(\rvbx;\rvbyr|\rvbye) = I(\rvbx;\rvbye,\rvbyr) - I(\rvbx;\rvbye)$,
and where the last inequality follows from the fact that
$\cov(\rvbz)=\bI$.
\end{IEEEproof}

\subsection{Property of the Saddle Point}
\label{sec:sadprop}

To simplify evaluation of the associated saddle value, we now develop
the Property~\ref{prop:degradedness}.  For notational convenience, we
define $\obLa$ via [cf.~\eqref{eq:bLa-def}]
\begin{equation} 
\obLa \defeq \bLa(\obK_P),\qquad \bLa(\bK_P) \defeq \bLa(\bK_P,\obK_\bPh).
\label{eq:obLa-def}
\end{equation}

The required property is obtained by combining the following two
lemmas.
\begin{lemma}
A saddle point solution $(\obK_P,\obK_\bPh)$ to \eqref{eq:capacityUB}
satisfies
\begin{equation} 
(\bHr -\obTh\bHe)\obK_P(\obPh^\dagger\bHr -\bHe)^\dagger = \bzero
\label{eq:KphCondn}
\end{equation}
\label{lem:saddlePointProperty-a}
\end{lemma}
\begin{lemma}
A saddle point solution $(\obK_P,\obK_\bPh)$ to \eqref{eq:capacityUB}
is such that
\begin{equation}
\text{$(\bHr-\obTh\bHe)\obS$ has a full column-rank}
\label{eq:KpCondn}
\end{equation} 
provided $\bHr\neq\obTh\bHe$, where $\obS$ is a full
column-rank matrix such that $\obS \obS^\dagger = \obK_P$.
\label{lem:saddlePointProperty-b}
\end{lemma}

In particular, combining \eqref{eq:KphCondn} and \eqref{eq:KpCondn} we
immediately obtain \eqref{eq:degradedness}, since for a full
column-rank matrix $\bM$, $\bM\ba = \bzero$ if and only if $\ba =
\bzero$.

In the remainder of the section, we prove the two lemmas.

\begin{IEEEproof}[Proof of Lemma~\ref{lem:saddlePointProperty-a}]
Here we consider the simpler case when $\obK_\bPh\succ\bzero$; the
extension of the proof to the case when $\obK_\bPh$ is singular is
provided in Appendix~\ref{app:KphCondnSing}.

We begin by noting that the second inequality
in \eqref{eq:saddlePointProp} implies
\begin{equation}
\obK_\bPh \in \argmin_{\bK_\bPh \in \cK_\bPh}\Rp(\obK_P,\bK_\bPh).
\label{eq:noiseMin}
\end{equation}
The Lagrangian associated with the minimization \eqref{eq:noiseMin} is
\begin{equation}
\cL_\bPh(\bK_\bPh, \bUp) = \Rp(\obK_P,\bK_\bPh) +
\tr(\bUp \bK_\bPh), 
\end{equation}
where the dual variable
\begin{equation}
\label{eq:UpsDefn}
\bUp= {\kbordermatrix{& \Nr & \Ne \cr 
                        \Nr & \bUp_1 & \bzero \\
\Ne  & \bzero & \bUp_2}}
\end{equation}
is a block diagonal matrix corresponding to the constraint that the
noise covariance $\bK_\bPh$ must have identity matrices on its
diagonal.  The associated KKT conditions yield
\begin{multline} 
\nabla_{\bK_\bPh}\cL_\bPh(\bK_\bPh, \bUp)\bigr|_{\bK_\bPh=\obK_\bPh} \\
=
\nabla_{\bK_\bPh}\Rp(\obK_P,\bK_\bPh)\bigr|_{\bK_\bPh=\obK_\bPh} + \bUp = 
  \bzero.
\label{eq:nablaEqn}
\end{multline}
Substituting
\begin{align}
&\nabla_{\bK_\bPh} \Rp(\obK_P,\bK_\bPh)\bigr|_{\bK_\bPh=\obK_\bPh} \notag\\
&\ = \nabla_{\bK_\bPh}\left[\log\det(\bK_\bPh +\bHt\obK_P\bHt^\dagger)\!-\!\log\det(\bK_\bPh)\right]\bigr|_{\bK_\bPh=\obK_\bPh}\notag\\
&\ =(\obK_\bPh + \bHt\obK_P\bHt^\dagger)^{-1} - \obK_\bPh^{-1},
\label{eq:noiseKKTCondn}
\end{align}
with \eqref{eq:bHt-def} into \eqref{eq:nablaEqn} and simplifying, we
obtain,
\begin{equation}
\bHt\obK_P\bHt^\dagger = \obK_\bPh\bUp(\obK_\bPh + \bHt\obK_P\bHt^\dagger).
\label{eq:noiseKKTCondnb}
\end{equation}

To complete the proof requires a straightforward manipulation of
\eqref{eq:noiseKKTCondnb} to obtain \eqref{eq:KphCondn}.
Specifically, substituting for $\obK_\bPh$ from \eqref{eq:oPh-def} and
$\bHt$ from \eqref{eq:bHt-def} into \eqref{eq:noiseKKTCondnb}, and
carrying out the associated block matrix multiplication yields
\begin{align}
\bHr \obK_P\bHr^\dagger &= \bUp_1(\bI + \bHr\obK_P\bHr^\dagger) +
\obPh\bUp_2(\obPh^\dagger + \bHe\obK_P\bHr^\dagger) \label{eq:Hrr}\\
\bHr\obK_P\bHe^\dagger &= \bUp_1(\obPh +
\bHr\obK_P\bHe^\dagger)+\obPh\bUp_2(\bI +
\bHe\obK_P\bHe^\dagger) \label{eq:Hre} \\ 
\bHe\obK_P\bHr^\dagger &= \obPh^\dagger\bUp_1(\bI +
\bHr\obK_P\bHr^\dagger)+ \bUp_2(\obPh^\dagger
+\bHe\obK_P\bHr^\dagger) \label{eq:Her} \\ 
\bHe\obK_P\bHe^\dagger &=\obPh^\dagger\bUp_1(\obPh +
\bHr\obK_P\bHe^\dagger) + \bUp_2(\bI + \bHe\obK_P\bHe^\dagger). \label{eq:Hee}
\end{align}
Eliminating $\bUp_1$ from \eqref{eq:Hrr} and \eqref{eq:Her}, we obtain
\begin{equation}
(\obPh^\dagger \bHr-\bHe)\obK_P\bHr^\dagger = (\obPh^\dagger\obPh -\bI)\bUp_2(\obPh^\dagger +\bHe\obK_P\bHr^\dagger),
\label{eq:ups21}
\end{equation}
and eliminating $\bUp_1$ from \eqref{eq:Hre} and \eqref{eq:Hee}, we obtain
\begin{equation}
(\obPh^\dagger\bHr-\bHe)\obK_P\bHe^\dagger = (\obPh^\dagger\obPh-\bI)\bUp_2(\bI + \bHe\obK_P\bHe^\dagger).
\label{eq:ups22}
\end{equation}
Finally, eliminating $\bUp_2$ from \eqref{eq:ups21} and
\eqref{eq:ups22}, we obtain
\begin{align}
&(\obPh^\dagger \bHr-\bHe)\obK_P\bHr^\dagger \notag\\
&= (\obPh^\dagger\bHr-\bHe)\obK_P\bHe^\dagger(\bI +
  \bHe\obK_P\bHe^\dagger)^{-1}(\obPh^\dagger
  +\bHe\obK_P\bHr^\dagger)\notag\\ 
&= (\obPh^\dagger\bHr-\bHe)\obK_P\bHe^\dagger\obTh^\dagger, \label{eq:elim-end}
\end{align}
which reduces to \eqref{eq:KphCondn} as desired.
\end{IEEEproof}

In preparation for proving Lemma~\ref{lem:saddlePointProperty-b}, we
establish the following key proposition, whose proof is provided in
Appendix~\ref{app:KKTOptimalityProof}.
\begin{prop}
\label{claim:HSaddlePoint}
When $\rvbx\sim\CN(\bzero,\bK_P)$ and $\rvbz\sim\CN(\bzero,\obK_\bPh)$ with
$\obK_\bPh\succ \bzero$ in the model \eqref{eq:channel}, we
have\footnote{Note that the maximum on the left-hand side is in
general a lower bound on the maximum on the right-hand side.}
\begin{equation}
\argmax_{\bK_P \in \cK_P} h(\rvbyr-\bTh(\bK_P)\rvbye)
= \argmax_{\bK_P \in \cK_P} h(\rvbyr-\obTh\rvbye),
\label{eq:secondOpt}
\end{equation}
where $\obTh$ and $\bTh(\bK_P)$ are as defined in \eqref{eq:obTh-def}
with \eqref{eq:bTh-def}.
\end{prop}

\begin{IEEEproof}[Proof of Lemma~\ref{lem:saddlePointProperty-b}]
Again, here we consider the simpler case when $\obK_\bPh$ is
nonsingular; a proof for the case when $\obK_\bPh$ is singular is
provided in Appendix~\ref{app:KphSingRankCondn}.

We begin by noting that
\begin{align}
\obK_P &\in \argmax_{\bK_P \in
  \cK_P}\Rp(\bK_P,\obK_\bPh) \label{eq:use-saddle-first} \\  
&=\argmax_{\bK_P \in \cK_P} h(\rvbyr|\rvbye) \label{eq:ent-rel}\\ 
&=\argmax_{\bK_P \in \cK_P} h(\rvbyr - \bTh(\bK_P) \rvbye)
\\ %\label{eq:mmseCondn1}
&=\argmax_{\bK_P \in \cK_P} h(\rvbyr - \obTh \rvbye)
\label{eq:use-prop} \\
&= \argmax_{\bK_P\in \cK_P} \log\det(\bI + \obHeff \bK_P
\obHeff^\dagger), \label{eq:KpMax2}  
\end{align}
where \eqref{eq:use-saddle-first} follows from the first inequality in
\eqref{eq:saddlePointProp}, where \eqref{eq:ent-rel} follows from the
fact that $\obK_\bPh\succ\bzero$, where \eqref{eq:use-prop} follows
from Proposition~\ref{claim:HSaddlePoint}, and where in
\eqref{eq:KpMax2} we have the effective channel\footnote{As an aside,
note that \eqref{eq:KpMax2} provides the interpretation of $\obK_P$ as
an optimal input covariance for a MIMO channel with matrix $\obHeff$
and unit-variance white Gaussian noise.}
\begin{subequations} 
\begin{equation}
\obHeff \defeq \obJ^{-1/2}(\bHr-\obTh\bHe),
\end{equation}
with
\begin{equation} 
\begin{aligned}
\obJ 
&\defeq \bI + \obTh\obTh^\dagger-\obTh\obPh^\dagger-\obPh\obTh^\dagger\\
&= (\bI-\obPh\obPh^\dagger) + (\obTh-\obPh)(\obTh-\obPh)^\dagger,
\end{aligned}%
\end{equation}%
\label{eq:bHeff-def}%
\end{subequations}
which is nonsingular since $\obK_\bPh\succ\bzero$.

Finally, because $\obJ\succ\bzero$, showing \eqref{eq:KpCondn} is
equivalent to showing that that $\obHeff\obS$ has full column-rank,
which we establish in the sequel to conclude the proof.  First, we
express $\obHeff$ in terms of its singular value decomposition
\begin{equation}
\obHeff = \bA \bSieff \bB^\dagger,
\label{eq:Heff-svd}
\end{equation}
i.e., $\bA$ and $\bB$ are unitary matrices, and
\begin{equation}
\bSieff = {\kbordermatrix{& \nu & \Nt-\nu \cr
                        \nu & \bSi_0 & \bzero \\
\Nr-\nu  & \bzero & \bzero}},
\label{eq:bSieff-def}
\end{equation}
where $\nu\defeq\rank(\obHeff)>0$ and $\bSi_0$ is diagonal with strictly
positive entries.  We establish that $\obHeff\obS$ has full
column-rank by showing that the columns of $\obS$ are spanned by the
first $\nu$ columns of $\bB$, i.e.,
\begin{equation}
\obF \defeq \bB^\dagger\obK_P \bB 
= \kbordermatrix{ & \nu & \Nt - \nu \\ \nu & \obF_0 & \bzero \\ \Nt-\nu
  & \bzero & \bzero}
\label{eq:Fstruct}
\end{equation}
for some $\obF_0\succeq\bzero$.

To this end, substituting \eqref{eq:Heff-svd} into
\eqref{eq:KpMax2}, we obtain
\begin{align}
\obK_P &= \argmax_{\bK_P\in\cK_P}\log\det(\bI + \bA\bSieff\bB^\dagger \bK_P\bB \bSieff^\dagger \bA^\dagger)\notag\\
&=\argmax_{\bK_P\in\cK_P}\log\det(\bI + \bSieff\bB^\dagger\bK_P\bB
\bSieff^\dagger), \label{eq:KpBMax}.
\end{align}
Now $\bK_P \in \cK_P$ if and only if
$\bF=\bB^\dagger\bK_P\bB\in\cK_P$, so \eqref{eq:KpBMax} implies that 
\begin{align}
\obF &\in \argmax_{\bF\in\cK_P}\log\det(\bI + \bSieff
\bF\bSieff^\dagger) \notag\\ %\label{eq:FMatDefn}
&= \argmax_{\bF\in\cK_P}\log\det(\bI + \bSi_0 \bF_0\bSi_0^\dagger),\label{eq:FmatDefn2}
\end{align}
with $\bF$ expressed in terms of the block notation
\begin{equation}
\bF = \kbordermatrix{ &
\nu & \Nt - \nu \\ \nu & \bF_0 & \bF_1 \\ \Nt-\nu &
\bF_1^\dagger & \bF_2}, \label{eq:Fstruct2}
\end{equation}
and where \eqref{eq:FmatDefn2} follows from \eqref{eq:bSieff-def}.

Finally, it follows that $\obF_1$ and $\obF_2$, the $\bF_1$ and
$\bF_2$ in \eqref{eq:Fstruct2} when $\bF=\obF$, are both $\bzero$.
Indeed, if $\obF_2 \neq 0$, then $\tr(\obF_2) > 0$.  This would
contradict the optimality in \eqref{eq:FmatDefn2}: since the objective
function only depends on $\obF_0$, one could strictly increase the
objective function by increasing the trace of $\obF_0$ and decreasing
the trace of $\obF_2$.  Finally, since $\obF \succeq \bzero$ and
$\obF_2 = 0$, it follows that $\obF_1 = 0$.
\end{IEEEproof}

\subsection{Evaluation of the Saddle Value: Proof of Theorem~\ref{thm1}}

The conditions in Lemmas~\ref{lem:saddlePointProperty-a} and
\ref{lem:saddlePointProperty-b} can be used in turn to establish the
tightness of the upper bound \eqref{eq:capacityUB}.

\begin{lemma}
The saddle value $\Rub$ in \eqref{eq:capacityUB} can be expressed
as
\begin{equation}
\Rub =
\begin{cases}
\Rm(\obK_P), & \bHr\neq\obTh\bHe \\
0, & \text{otherwise},
\end{cases}
\label{eq:UpperBoundSimplified}
\end{equation}
where $\Rm(\obK_P)$ is as given in \eqref{eq:Rm-def}.
\label{lem:upperBoundSimplified}
\end{lemma}

The proof of Theorem~\ref{thm1} is a direct consequence of
Lemma~\ref{lem:upperBoundSimplified}.  If $\Rp(\obK_P,\obK_\bPh)=0$,
the capacity is zero, otherwise $\Rp(\obK_P,\obK_\bPh)=\Rm(\obK_P)$,
and the latter expression is an achievable rate as can be seen by
setting $p_\rvu = p_\rvbx=\CN(\bzero,\obK_P)$ in the argument of
\eqref{eq:CK}.

Thus, to conclude the section it remains only to prove our lemma.
\begin{IEEEproof}[Proof of Lemma~\ref{lem:upperBoundSimplified}]
Here we consider the case when when $\obK_\bPh\succ\bzero$, i.e.,
$\|\bPh\|_2 < 1$; the proof for the case when $\obK_\bPh$ is singular
is provided in Appendix~\ref{app:upper_Bound_Simplified_Singular}.

To obtain \eqref{eq:UpperBoundSimplified} when $\bHr \neq \obTh\bHe$,
we begin by writing the gap between upper and lower bounds as
\begin{align}
\Rp&(\obK_P,\obK_\bPh) - \Rm(\obK_P) \notag\\
&\quad=  I(\rvbx;\rvbyr|\rvbye)- \left[
  I(\rvbx;\rvbyr)-I(\rvbx;\rvbye) \right] \notag\\ 
&\quad= I(\rvbx;\rvbye|\rvbyr) \label{eq:I-gap}\\ 
&\quad= h(\rvbye|\rvbyr)-h(\rvbze|\rvbzr), \notag
\end{align}
then note that this gap is zero since
\begin{align}
&h(\rvbye|\rvbyr) \notag\\
&= \log \det\pi e \bLab \label{eq:sub-mmset}\\
&= \log\det\pi e (\bI+\bHe\obK_P\bHe^\dagger-\obPh^\dagger(\bI+\bHr\obK_P\bHr^\dagger)\obPh) \label{eq:use-nc-1}\\  
&=\log\det\pi e(\bI-\obPh^\dagger\obPh) \label{eq:use-nc-2}\\
&= h(\rvbze|\rvbzr),\notag  %\label{eq:Leq3}
\end{align}
where in \eqref{eq:sub-mmset}
\begin{align} 
&\bLab = \notag\\
&\ \ \bI+\bHe\obK_P\bHe^\dagger \notag\\
&\quad {} - (\obPh^\dagger + \bHe\obK_P\bHr^\dagger) 
(\bI + \bHr\obK_P\bHr^\dagger)^{-1} (\obPh+\bHr\obK_P\bHe^\dagger))
\end{align}
is the ``backward'' error covariance associated with the linear MMSE
estimate of $\rvbye$ from $\rvbyr$, and where to obtain each of
\eqref{eq:use-nc-1} and \eqref{eq:use-nc-2} we have used
\eqref{eq:degradedness} of Property~\ref{prop:degradedness}.

To obtain \eqref{eq:UpperBoundSimplified} when $\bHr=\obTh\bHe$, we
note that
\begin{align}
\Rp(\obK_P,\obK_\bPh) &= I(\rvbx;\rvbyr|\rvbye)\label{eq:HrHecond0}\\
&= h(\rvbyr|\rvbye)-h(\rvbzr|\rvbze) \notag\\
&= h(\rvbyr - \obTh\rvbye) - h(\rvbzr-\obPh\rvbze)\label{eq:mmseStep}\\
&= h(\rvbzr - \obTh\rvbze) - h(\rvbzr-\obPh\rvbze)\label{eq:HrHecond}\\
&=0, \label{eq:th-ph}
\end{align}
where \eqref{eq:mmseStep} follows from the fact that $\obTh$ in
\eqref{eq:obTh-def} is the coefficient in the MMSE estimate of
$\rvbyr$ from $\rvbye$, and $\obPh$ is the coefficient in the MMSE
estimate of $\rvbzr$ from $\rvbze$, where \eqref{eq:HrHecond} follows
via the relation $\bHr=\obTh\bHe$, so that $\rvbyr - \obTh\rvbye =
\rvbzr - \obTh\rvbze$, and where \eqref{eq:th-ph} follows from
\eqref{eq:zero-cap-obTh}.
\end{IEEEproof}

\section{Capacity Analysis in the High-SNR Regime}
\label{sec:HighSNR}

We begin with a convenient upper bound that is used in our converse
argument, then exploit the GSVD in developing the coding scheme for
our achievability argument.  Our high-SNR capacity results follow, and
separately consider the cases where $\bHe$ does and does not have full
column-rank.

\begin{lemma}
\label{lem:c-ub}
For all choices of $\bTh \in \compls^{\Nr \times \Nt}$ and
$\bPh\in\compls^{\Nr\times\Ne}$ such that $\|\bPh\|_2 \le 1$, the
secrecy capacity \eqref{eq:capacityUB} of the channel
\eqref{eq:channel} is upper bounded by
\begin{subequations} 
\begin{equation}
C(P) \le \max_{\obK_P \in\cK_P}\Rpp(\bK_P,\bTh,\bPh),
\label{eq:c-ub}
\end{equation}
where
\begin{align}
&\Rpp(\bK_P,\bTh,\bPh) \notag\\
&\quad\defeq h(\rvbyr-\bTh\rvbye)- \log\det\pi e(\bI-\bPh\bPh^\dagger) \notag\\
&\quad= \log \frac{\det(\bHh\bK_P\bHh^\dagger + \bI +
  \bTh\bTh^\dagger - \bTh\bPh^\dagger -
  \bPh\bTh^\dagger)}{\det(\bI-\bPh\bPh^\dagger)}, 
\label{eq:Rpp-def}
\end{align}
with 
\begin{equation} 
\bHh = \bHr - \bTh \bHe.
\label{eq:Heff-def}
\end{equation}%
\label{eq:AlternateUB}%
\end{subequations}
\end{lemma}

\begin{IEEEproof}
First note that the objective function $\Rp(\bK_P,\bK_\bPh)$ in
\eqref{eq:Rplus-def} can be expressed in the form
\begin{align}
\Rp(\bK_P,\bK_\bPh)
 &= I(\rvbx;\rvbyr|\rvbye)\notag\\
&= h(\rvbyr|\rvbye)- h(\rvbzr|\rvbze) \notag\\
&= h(\rvbyr|\rvbye)- \log\det\pi e(\bI-\bPh\bPh^\dagger) \notag\\
&= \min_{\bTh} h(\rvbyr-\bTh\rvbye)- \log\det\pi
 e(\bI-\bPh\bPh^\dagger) \notag\\
&= \min_{\bTh} \Rpp(\bK_P,\bTh,\bPh), \label{eq:use-Rpp}
\end{align}
Hence,
\begin{align}
\Rub &= \min_{\cK_\bPh}\max_{\cK_P}
\Rp(\bK_P,\bK_\bPh) \label{eq:Rub-a} \\
&= \min_{\cK_\bPh}\max_{\cK_P} \min_{\bTh}
\Rpp(\bK_P,\bTh,\bPh) \label{eq:Rub-b} \\ 
&\le \min_{\cK_\bPh}\min_{\bTh}\max_{\cK_P}
\Rpp(\bK_P,\bTh,\bPh) \label{eq:Rub-c} \\
& = \min_{\bPh:\|\bPh\|_2\le 1}\min_{\bTh}\max_{\bK_P\in\cK_P}
\Rpp(\bK_P,\bTh,\bPh), 
\label{eq:minminmax}
\end{align}
where to obtain \eqref{eq:Rub-a} we have used \eqref{eq:capacityUB},
where to obtain \eqref{eq:Rub-b} we have used \eqref{eq:use-Rpp}, and
where to obtain \eqref{eq:Rub-c} we have used that a minimax quantity
upper bounds a corresponding maximin quantity.

Finally, we further upper bound \eqref{eq:minminmax} by making
arbitrary choices for $\bTh$ and $\bPh$, yielding
\eqref{eq:AlternateUB}.
\end{IEEEproof}

\subsection{GSVD Properties}
\label{sec:GSVD_defn}

The following properties of the GSVD in Definition~\ref{defn:gsvd} are
useful in our analysis.

First, the GSVD simultaneously diagonalizes the channels in our model
\eqref{eq:channel}.    In particular, applying \eqref{eq:gsvd-def} we
obtain
\begin{equation} 
\begin{aligned}
\tilde{\rvby}_\mrm{r}(t) &= \tilde{\bSi}_\mrm{r} \tilde{\rvbx}(t) +
\tilde{\rvbz}_\mrm{r}(t)\\ 
\tilde{\rvby}_\mrm{e}(t) &= \tilde{\bSi}_\mrm{e} \tilde{\rvbx}(t) +
\tilde{\rvbz}_\mrm{e}(t), 
\end{aligned}
\label{eq:channel-diag}
\end{equation}
where
\begin{align*} 
\tilde{\bSi}_\mrm{r} &= {\kbordermatrix{& k-p-s & s & p\\
                         s & \bzero & \bDr & \bzero \\
                         p  & \bzero & \bzero & \bI}} \\
\tilde{\bSi}_\mrm{e} &= {\kbordermatrix{& k-p-s & s & p\\
                         k-p-s & \bI & \bzero & \bzero \\
                         s  & \bzero & \bDev & \bzero}}.
\end{align*}
and
\begin{align*} 
\tilde{\rvbx}(t) &= \bOm^{-1} \bigl[\bPsi_\mrm{t}^\dagger
  \rvbx(t)\bigr]_{1:k} \\
\tilde{\rvby}_\mrm{r}(t) &= \bigl[ \bPsi_\mrm{r}^\dagger \rvby_\mrm{r}(t)
  \bigr]_{n_\mrm{r}-p-s+1:n_\mrm{r}} \\
\tilde{\rvby}_\mrm{e}(t) &= \bigl[ \bPsi_\mrm{e}^\dagger \rvby_\mrm{e}(t)
  \bigr]_{1:k-p} \\
\tilde{\rvbz}_\mrm{r}(t) &= \bigl[ \bPsi_\mrm{r}^\dagger \rvbz_\mrm{r}(t)
  \bigr]_{n_\mrm{r}-p-s+1:n_\mrm{r}} \\
\tilde{\rvbz}_\mrm{e}(t) &= \bigl[ \bPsi_\mrm{e}^\dagger \rvbz_\mrm{e}(t)
  \bigr]_{1:k-p}.
\end{align*}

\begin{figure}[tbp]
\psfrag{&x}{$\tilde{\rvbx} \quad \left\{ \rule[-.43755in]{0in}{.875in} \right.$}
\psfrag{&ye}{$\left.  \rule[-.21875in]{0in}{.4375in} \right\} \quad
  \tilde{\rvby}_\mrm{e}$} 
\psfrag{&yr}{$\left.\rule[-.21875in]{0in}{.4375in} \right\} \quad
  \tilde{\rvby}_\mrm{r}$} 
\psfrag{&ze}{$\,\tilde{\rvbz}_\mrm{e}$}
\psfrag{&zr}{$\,\tilde{\rvbz}_\mrm{r}$}
\psfrag{&1}{{\footnotesize $k-p-s$}}
\psfrag{&2}{{\footnotesize $s$}}
\psfrag{&3}{{\footnotesize $p$}}
\psfrag{&r}{$r_{1:s}$}
\psfrag{&e}{$e_{1:s}$}
\includegraphics[width=2.75in]{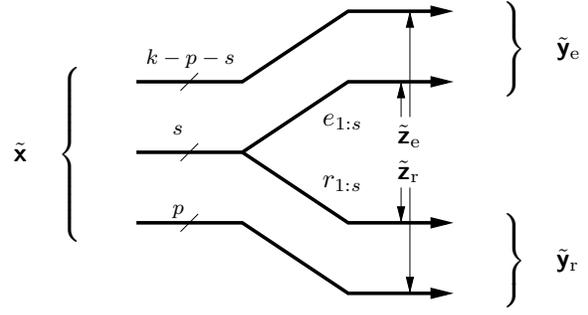}
\caption{Equivalent parallel channel model obtained via GSVD.  \label{fig:diag}}
\end{figure}

The corresponding equivalent channel is as depicted in
Fig.~\ref{fig:diag}.

Second, the GSVD yields a characterization of the null space of $\bHe$.
In particular, 
\begin{equation} 
\Null(\bHe)=\cSr\cup\cSn,
\label{eq:null-union}
\end{equation}
where, expressing $\bPsit$ as defined in \eqref{eq:gsvd-def} in terms
of its columns $\bps_i$, $i=1,\dots,\Nt$, \viz,
\begin{equation*}
\bPsit = \begin{bmatrix} \bpsi_1 & \cdots & \bpsi_{\Nt} \end{bmatrix},
\end{equation*}
we have [cf.~\eqref{eq:cSr-def}, \eqref{eq:cSn-def}]
\iffalse
\footnote{Although not required in our
  analysis, we also have
\begin{align*}
\cSre &= \spn(\bpsi_{k-p-s+1},\ldots, \bpsi_{k-p}) \\
\cSe &= \spn(\bpsi_{1},\ldots, \bpsi_{k-p-s}).
\end{align*}
}
\fi
\begin{subequations}
\begin{align}
\cSr &= \spn(\bpsi_{k-p+1},\ldots, \bpsi_{k})
\label{eq:cSr-basis} \\
\cSn &= \spn(\bpsi_{k+1},\ldots, \bpsi_{\Nt}).
\label{eq:cSn-basis}
\end{align}%
\label{eq:cS-bases}%
\end{subequations}

We first verify \eqref{eq:cS-bases}.  To establish
\eqref{eq:cSn-basis}, it suffices to note that
\begin{equation*}
\bHr\bpsi_j = \bHe\bpsi_j = \bzero, \quad j= k+1,\ldots, \Nt,
\end{equation*}
which can be readily verified from \eqref{eq:gsvd-def}.   

To establish \eqref{eq:cSr-basis}, we show for all $j\in\{k-p+1,
\dots, k\}$ that $\bHe\bpsi_j = \bzero$ and that the $\{\bHr \bpsi_j
\}$ are linearly independent.  It suffices to show that the last $p$
columns of $\bSir\bOm^{-1}$ are linearly independent and the last $p$
columns of $\bSie\bOm^{-1}$ are zero.  To this end, note that since
$\bOm^{-1}$ in \eqref{eq:gsvd-def} is a lower triangular matrix, it
can be expressed in the form
\begin{equation}
\bOm^{-1}= {\kbordermatrix{& k-p-s & s & p \cr 
           k-p-s & \bOm_{1}^{-1} & \bzero & \bzero \\ 
           s & \bT_{21} & \bOm_2^{-1} & \bzero \\ 
           p &\bT_{31} & \bT_{32} & \bOm_{3}^{-1}}}.
\label{eq:Omeg-def}
\end{equation}
By direct block left-multiplication of \eqref{eq:Omeg-def} with
\eqref{eq:Sir-defn} and \eqref{eq:Sie-defn}, we have
\begin{subequations}
\begin{equation} 
\bSir\bOm^{-1}= {\kbordermatrix{& k-p-s & s & p \cr 
k-p-s &  \bzero & \bzero & \bzero \\
s & \bDr\bT_{21} & \bDr\bOm_2^{-1} & \bzero \\ 
p & \bT_{31} & \bT_{32} & \bOm_{3}^{-1}}}
\label{eq:SirBlock}  \\
\end{equation}%
\begin{equation} 
\bSie\bOm^{-1} = {\kbordermatrix{& k-p-s & s & p \cr 
k-p-s & \bOm_1^{-1} & \bzero & \bzero \\ 
s & \bDev\bT_{21} & \bDev\bOm_{2}^{-1} & \bzero \\ 
p & \bzero & \bzero & \bzero}}.
\label{eq:SieBlock}%
\end{equation}%
\end{subequations}
Since $\bOm_3$ is invertible (since $\bOm$ is nonsingular), the last
$p$ columns of $\bSir\bOm^{-1}$ are linearly independent and the last
$p$ columns of $\bSie\bOm^{-1}$ are zero, establishing
\eqref{eq:cSr-basis}.

To characterize $\Null(\bHe)$, we use \eqref{eq:cSr-basis} and
\eqref{eq:cSn-basis} with \eqref{eq:null-union} to obtain
\begin{equation*}
\Null(\bHe)= \spn(\bpsi_{k-p+1}, \ldots, \bpsi_{\Nt}),
\end{equation*}
from which we obtain that 
\begin{equation}
\bHe^\sharp = \bPsi_\mrm{ne}\bPsi_\mrm{ne}^\dagger,
\label{eq:HePerpDefn}
\end{equation}
is the projection matrix onto $\Null(\bHe)$, where
\begin{equation} 
\bPsi_\mrm{ne} = 
\begin{bmatrix} 
\bpsi_{k-p+1} & \cdots & \bpsi_{\Nt} 
\end{bmatrix}.
\label{eq:ne-def}
\end{equation}
In turn, using \eqref{eq:ne-def} and \eqref{eq:SirBlock} in
\eqref{eq:HrGsvd} we obtain
\begin{equation*}
\bHr{\bPsi}_\mrm{ne}
=\bPsir \left\{\kbordermatrix{  & p & \Nt-k \cr \Nr-p & \bzero & \bzero\\ p &
  \bOm_3^{-1} & \bzero}\right\}, 
\end{equation*}
whence
\begin{equation}
\bHr\bHe^\sharp\bHr^\dagger 
= \bPsir\left\{ \kbordermatrix{ & \Nr-p & p \cr 
                \Nr-p & \bzero & \bzero & \\ 
                 p & \bzero &\bOm_3^{-1}\bOm_3^{-\dag}}
         \right\} \bPsir^\dagger.
\label{eq:HrHe_perp}
\end{equation}

Third, the GSVD can be more simply described when the matrix
$\bHe$ has a full column-rank.  To see this, first note from
\eqref{eq:k-def} and \eqref{eq:p-s-def} that 
\begin{equation} 
k=\Nt \quad\text{and}\quad p=0,
\label{eq:kp-fullrank}
\end{equation}
respectively, and thus \eqref{eq:gsvd-def} specializes to
\begin{subequations}
\begin{equation}
\bPsir^\dagger \bHr \bPsit \bOm = \bSir,\quad \bPsie^\dagger \bHe \bPsit\bOm = \bSie,
%\label{eq:GsvdDefn_FullRank}
\end{equation}
with [cf.~\eqref{eq:Sire-defn}]
\begin{equation}
\bSir = {\kbordermatrix{& \Nt-s & s \cr
                        \Nr-s & \bzero &  \bzero \\
                        s & \bzero & \bDr}},\quad
\bSie = {\kbordermatrix{& \Nt-s & s \cr
                        \Nt-s & \bI & \bzero \\
                        s  & \bzero & \bDev \\
                        \Ne-\Nt & \bzero & \bzero}},
%\label{eq:SiSeDefn_FullRank}%
\end{equation}%
\label{eq:gsvd-fullrank}%
\end{subequations}
and $\bDr$ and $\bDev$ as in \eqref{eq:Dre-defns}.
Hence, it follows from \eqref{eq:gsvd-fullrank} that
\begin{equation}
\bHe^\ddag \defeq \bPsit\bOm
\left\{
\kbordermatrix{ & \Nt-s & s & \Ne-\Nt \cr 
                \Nt-s & \bI & \bzero & \bzero\\ 
                s & \bzero & \bDev^{-1} & \bzero}
\right\} \bPsie^\dag
\label{eq:Heinv}
\end{equation}
satisfies $\bHe^\ddag \bHe=\bI$ and thus is the Moore-Penrose
pseudo-inverse of $\bHe$.  Finally, from \eqref{eq:gsvd-fullrank} and
\eqref{eq:Heinv} we obtain
\begin{equation*}
\bHr\bHe^\ddag = \bPsir 
\left\{
\kbordermatrix{
& \Nt-s &s&
\Ne - \Nt \\ 
\Nr - s & \bzero & \bzero &\bzero \\
s & \bzero & \bDr\bDev^{-1}& \bzero}\right\}\bPsie^\dagger,
%\label{eq:HrHeDecomp}
\end{equation*}
from which we see that the generalized singular values of ($\bHr$,
$\bHe$) in \eqref{eq:GSV-defn} are also the (ordinary) singular values
of $\bHr\bHe^\ddag$.

We now turn to our secrecy capacity analysis in the high-SNR regime.
There are two cases, which we consider separately.

\subsection{Case I: $\rank(\bHe)=\Nt$}
\label{subsec:GSVD_Cap} 

In this case, we use that \eqref{eq:kp-fullrank} holds and so the GSVD
is given by \eqref{eq:gsvd-fullrank}, and thus $\dim\cSre=s$,
$\dim\cSe=\Nt-s$, and $\dim\cSr=\dim\cSn=0$.

\subsubsection*{Achievability}

In the equivalent parallel channel model of Fig.~\ref{fig:diag}, there
are $s$ subchannels that go to the intended receiver (and also to the
eavesdropper, with different gains), which correspond to $\cSre$.  Of
these $s$ subchannels, we use only the subset for which the gains to
the intended receiver are stronger than those to the eavesdropper, and
with these our communication scheme uses Gaussian wiretap codebooks.

In particular, we transmit
\begin{equation}
\rvbx = \bPsit\bOm \begin{bmatrix} \bzero_{\Nt-s} \\ 
\rvbu\end{bmatrix}, \quad \rvbu =
[0,\ldots,0,\rvu_{\nu},\rvu_{\nu+1},\ldots,\rvu_s],
\label{eq:achievParams}
\end{equation}
where $\nu$ is the smallest integer such that $\sigma_j > 1$, and
where the nonzero elements of $\rvbu$ are i.i.d.\ $\CN(0, \al P)$ with
$\al = 1/(\Nt\sigma_{\max}(\bOm))$ so that the transmitted
power is at most $P$.

Using \eqref{eq:achievParams} and \eqref{eq:gsvd-fullrank} in
\eqref{eq:channel}, the observations at the intended receiver and
eavesdropper, respectively, take the form
\begin{equation*} 
\rvbyr = \bPsir  \begin{bmatrix}\bzero_{\Nt-s} \\\bDr
  \rvbu\end{bmatrix} + \rvbzr, 
\quad  \rvbye = \bPsie
  \begin{bmatrix}\bzero_{\Nt-s} \\ \bDev\rvbu \\
    \bzero_{\Ne-\Nt} \end{bmatrix} + \rvbze.
%\label{eq:ParalleModel} 
\end{equation*}
In turn, via \eqref{eq:CK}, the (secrecy) rate achievable with this
system is
\begin{align*}
%\label{eq:RachievStart}
R&= I(\rvbu;\rvbyr)-I(\rvbu;\rvbye) \\ 
&= \sum_{j=\nu}^{\Nt}\log\frac{1  + \al P r_j^2}{1 + \al P e_j^2}  \\
&= \sum_{j: \sigma_j > 1} \log \sigma_j^2 - o(1) 
%\label{eq:RachievStop}
\end{align*}
as required.
\hfill\IEEEQEDclosed

\subsubsection*{Converse}

It suffices to use Lemma~\ref{lem:c-ub} with the choices
\begin{subequations} 
\begin{equation}
\bTh = \bHr\bHe^\ddag, 
\quad \bPh = \bPsir\left\{
\kbordermatrix{& \Nt-s & s & \Ne-\Nt \\ 
\Nr-s & \bzero & \bzero &\bzero \\
s & \bzero & \bDepar & \bzero}
\right\}\bPsie^\dagger,
\label{eq:FullRankParams}
\end{equation}
where
\begin{equation}
\bDepar = \diag(\depar_1,\depar_2,\ldots,\depar_s),
\qquad \depar_i = \min\left(\sigma_i, \frac{1}{\sigma_i}\right),
\label{eq:Delta_Def}%
\end{equation}%
\end{subequations}
and where $\bHe^\ddag$ is the pseudo-inverse defined in
\eqref{eq:Heinv}.  With these choices of parameters,
\eqref{eq:Heff-def} evaluates to $\bHh=\bzero$, so we can ignore the
maximization over $\bK_P$ in \eqref{eq:c-ub}.  Simplifying
\eqref{eq:AlternateUB} for our choice of parameters yields
\begin{align}
\Rpp &\le \log\frac{\det(\bI+ (\bDr\bDev^{-1})^2-2 \bDr\bDev^{-1}\bDepar)}{\det(\bI - \bDepar^2)} \notag\\
&= \sum_{j:\sigma_j>1}\log\sigma_j^2,
\label{eq:bnd-simp}
\end{align}
which establishes our result.
\hfill\IEEEQEDclosed

\subsection{Case II: $\rank(\bHe)<\Nt$}

In this case, we use the general form of the GSVD as given by
\eqref{eq:gsvd-def}, so now $\dim\cSr=p>0$ and $\dim\cSre=s>0$.

\subsubsection*{Achievability}

In the equivalent parallel channel model of Fig.~\ref{fig:diag}, there
are $p$ subchannels that go only to the intended receiver,
corresponding to $\cSr$, and $s$ subchannels that go to both the
intended receiver and eavesdropper (with different gains),
corresponding to $\cSre$.  Our communication scheme uses both sets of
subchannels independently with Gaussian (wiretap) codebooks.  

In particular, we transmit
\begin{equation}
\rvbx = \bPsit 
\begin{bmatrix} 
\bzero_{k-p-s}\\ \bOm_{2}\rvbu \\ \rvbv \\ \bzero_{\Nt-k} 
\end{bmatrix},
\label{eq:genCaseParams}
\end{equation}
where $\rvbv$ and $\rvbu$ are the length-$p$ and length-$s$ auxiliary
random vectors associated with communication over $\cSr$ and
$\cSre$, respectively.  The elements of $\rvbv$ are i.i.d.\
$\CN(0,(P-\sqrt{P})/p)$, corresponding to allocating power
$P-\sqrt{P}$ to $\cSr$.  For $\cSre$, we use only the
subset of channels for which the gains to the intended receiver are
stronger than those to the eavesdropper, so $\rvbu = [0,\ldots, 0,
\rvu_{\nu},\ldots, \rvu_{s}]^\T$, where $\nu$ is the smallest integer
such that $\sigma_j > 1$, and where the nonzero elements are i.i.d.\
$\CN(0, \al\sqrt{P})$, independent of $\rvbv$, with $\al =
1/(\Nt\sigma_{\max}(\bOm_{2}))$ so that the power allocated to
$\cSre$ is at most $\sqrt{P}$.

With $\rvbx$ as in \eqref{eq:genCaseParams}, the observations at the
intended receiver and eavesdropper, respectively, take the form
\begin{subequations}
\begin{align} 
\rvbyr &= \bPsir 
\begin{bmatrix} \bzero_{\Nr-p-s} \\ \bDr\rvbu
  \\\bT_{32}\bOm_{2}^{-1}\rvbu + \bOm_3^{-1}\rvbv \end{bmatrix} 
+ \rvbzr, \label{eq:simplifiedYeqns-r} \\
\rvbye &= \bPsie 
\begin{bmatrix}\bzero_{k-p-s} \\ \bDev \rvbu \\
  \bzero_{\Ne+p-k} \end{bmatrix} +\rvbze. \label{eq:simplifiedYeqns-e}
\end{align}%
\label{eq:simplifiedYeqns}%
\end{subequations}

Via \eqref{eq:CK}, the system \eqref{eq:simplifiedYeqns} achieves
(secrecy) rate
\begin{align}
R &= I(\rvbu,\rvbv;\rvbyr)-I(\rvbu,\rvbv;\rvbye)\notag\\
&= I(\rvbu;\rvbyr)-I(\rvbu;\rvbye) +I(\rvbv;\rvbyr|\rvbu),
\label{eq:use-indep}
\end{align}
where \eqref{eq:use-indep} follows from the fact that $\rvbv$ is
independent of $(\rvbye,\rvbu)$, as \eqref{eq:simplifiedYeqns-e}
reflects.  

Evaluating the terms in \eqref{eq:use-indep}, we obtain
\begin{align} 
I(\rvbu;\rvbyr)-I(\rvbu;\rvbye) 
&= \sum_{j=\nu}^{\Nt}\log\frac{1+\al\sqrt{P} r_j^2}{1 + \al \sqrt{P}
  e_j^2} \notag \\ 
&= \sum_{j: \sigma_j > 1} \log \sigma_j^2 - o(1),
\label{eq:Rachiev1}
\end{align}
and 
\begin{align} 
I(\rvbv;\rvbyr|\rvbu) 
&=
\log\det\left(\bI+\frac{P-\sqrt{P}}{p}\bOm_3^{-1}\bOm_{3}^{-\dag}\right)
\notag\\
&=\log\det\left(\bI +\frac{P}{p}\bOm_3^{-1}\bOm_{3}^{-\dag}\right) -
o(1) \label{eq:contLog}\\ 
&= \log\det\left(\bI+\frac{P}{p}\bHr\bHe^\sharp\bHr^\dagger\right) -
o(1),
\label{eq:Rachiev2}
\end{align}
where \eqref{eq:contLog} follows from the continuity of
$\log\det(\cdot)$, and where \eqref{eq:Rachiev2} follows from
\eqref{eq:HrHe_perp}.  Substituting \eqref{eq:Rachiev1} and
\eqref{eq:Rachiev2} into \eqref{eq:use-indep} yields our desired result.
\hfill\IEEEQEDclosed

\subsubsection*{Converse}

\begin{figure*}[!t]
\normalsize
\setcounter{tempcount}{\value{equation}}
\setcounter{equation}{122}
\begin{multline} 
\bHh\bTh\bHh^\dagger + \bI + \bTh\bTh^\dagger - \bTh\bPh^\dagger - \bPh\bTh^\dagger \\
= \bPsir\left\{\kbordermatrix{ &\Nr-s-p & s& p \cr
\Nr-s-p & \bI & \bzero & \bzero \\
s & \bzero & \bI+ (\bDr\bDev^{-1})^2-2 \bDr\bDev^{-1}\bDepar & (\bDr\bDev^{-1}- \bDepar)\bF_{32}^\dagger \\
p & \bzero & \bF_{32}(\bDr\bDev^{-1}-\bDepar) & \bI + \bF_{31}\bF_{31}^\dagger + \bF_{32}\bF_{32}^\dagger + \bOm_3^{-1}\bQ \bOm_3^{-\dag}}\right\} \bPsir^\dagger
\label{eq:full_simplify}
\end{multline}
\hrulefill
\setcounter{equation}{\value{tempcount}}
%\vspace*{4pt}
\end{figure*}

To establish the converse, we use Lemma~\ref{lem:c-ub} with the
choices
\begin{equation}
\bTh = \bPsir \left\{\kbordermatrix{ & k-s-p & s & \Ne + p-k \cr
\Nr-s-p & \bzero & \bzero & \bzero \\ s & \bzero
&\bDr\bDev^{-1} & \bzero \\ p & \bF_{31} & \bF_{32} &
\bzero }\right\} \bPsie^\dagger
\label{eq:bThchoice_gen} 
\end{equation}
and
\begin{equation}
\bPh = \bPsir\left\{\kbordermatrix{ & k-s-p & s & \Ne + p-k \cr
\Nr-s-p & \bzero & \bzero & \bzero \\ s & \bzero & \bDepar & \bzero \\ p
& \bzero & \bzero & \bzero }\right\}  \bPsie^\dagger,
\label{eq:bPhchoice_gen}
\end{equation}
where $\bDepar$ is as defined in \eqref{eq:Delta_Def}, and where 
we choose
\begin{align*}
\bF_{32} &= \bT_{32}\bOm_2\bDev^{-1}\\
\bF_{31} &= (\bT_{31}-\bF_{32}\bDev\bT_{21})\bOm_1
\end{align*}
with $\bT_{21}$, $\bT_{31}$ and $\bT_{32}$ as defined
in \eqref{eq:Omeg-def}, so that
\begin{align*}
\bHr &- \bTh\bHe \notag\\
&= \bPsir \bigl( 
\begin{bmatrix}\bSir \bOm^{-1} & \bzero_{\Nr\times\Nt-k}\end{bmatrix} \notag\\ 
&\quad\quad {}- \bPsir^\dagger \bTh \bPsie 
\begin{bmatrix}\bSie \bOm^{-1} & \bzero_{\Ne \times \Nt-k} \end{bmatrix}
\bigr) \bPsit^\dagger\\
&=\bPsir\left\{\kbordermatrix{&k-p-s & s & p & \Nt - k \cr
\Nr-s-p &  \bzero & \bzero & \bzero & \bzero \\
s & \bzero & \bzero & \bzero & \bzero \\
p & \bzero & \bzero & \bOm_3^{-1} & \bzero}\right\} \bPsit^\dagger.
\end{align*}
The upper bound expression \eqref{eq:AlternateUB} can now be
simplified as follows.
\begin{align}
&\bHh\bK_P \bHh^\dagger = (\bHr - \bTh\bHe)\bK_P(\bHr
  - \bTh\bHe)^\dagger \notag\\ 
&\quad = \bPsir \left\{\kbordermatrix{ & \Nr-p-s & s & p \\ \Nr-p-s & \bzero
    & \bzero & \bzero \\ s & \bzero &\bzero & \bzero \\ p & \bzero &
    \bzero &\bOm_3^{-1}\bQ\bOm_3^{-\dag}} \right\} \bPsir^\dagger,\label{eq:Heff_Simplified}
\end{align}
where $\bQ$ is related to $\bK_P$ via
\iffalse
\begin{equation*} 
\bigl[ \bPsit^\dagger\bK_P\bPsi_t \bigr]_{k-p+1:k+1,k-p+1:k+1} = \bQ,
\end{equation*}
i.e., 
\fi
\begin{equation*}
\bPsit^\dagger\bK_P\bPsi_t = \kbordermatrix{& k-p & p & \Nt - k \cr
k-p & & & \\
p  &  & \bQ &  \\
\Nt-k & & & },
\end{equation*}
and satisfies $\tr(\bQ) \le P$.
From \eqref{eq:Heff_Simplified}, \eqref{eq:bPhchoice_gen}
and \eqref{eq:bThchoice_gen}, we have that the numerator in the
right-hand side of \eqref{eq:Rpp-def} simplifies to
\eqref{eq:full_simplify} at the top of the next page.
\addtocounter{equation}{1}
% moved this figure definition to beginning of converse section

In turn, using \eqref{eq:full_simplify} and the Fischer inequality
(which generalizes Hadamard's inequality) for positive semidefinite
matrices \cite{bapat97}, we obtain
\begin{equation*}
\begin{aligned}
&\log\det(\bI + \bHh\bTh\bHh^\dagger + \bTh\bTh^\dagger - \bTh\bPh^\dagger - \bPh\bTh^\dagger)\\
&\quad \le  \log\det(\bI+ (\bDr\bDev^{-1})^2-2
  \bDr\bDev^{-1}\bDepar) \\ 
&\qquad\quad + \log\det(\bI + \bF_{31}\bF_{31}^\dagger + \bF_{32}\bF_{32}^\dagger + \bOm_3^{-1}\bQ \bOm_3^{-\dag}),
\end{aligned}
%\label{eq:logdetbnd}
\end{equation*}
which when used with \eqref{eq:AlternateUB} yields
\begin{align*}
&C(P)\\
&\ \le \log\frac{\det(\bI+ (\bDr\bDev^{-1})^2-2
  \bDr\bDev^{-1}\bDepar)}{\det(\bI - \bDepar^2)}\\ 
&\ \quad{}+ \!\! \max_{\substack{\bQ\succeq \bzero: \\ \tr(\bQ)\le P}}
\log\det(\bI + \bF_{31}\bF_{31}^\dagger + \bF_{32}\bF_{32}^\dagger +
\bOm_3^{-1}\bQ \bOm_3^{-\dag}), 
%\label{eq:full_simplify_2}
\end{align*}
the first term of which is identical to \eqref{eq:bnd-simp}.  Thus, it
remains only to establish that
\begin{align}
\max_{\substack{\bQ\succeq \bzero: \\ \tr(\bQ)\le P}} 
&\log\det(\bI + \bF_{31}\bF_{31}^\dagger + \bF_{32}\bF_{32}^\dagger +
\bOm_3^{-1}\bQ \bOm_3^{-\dag})\notag\\ 
&\le \log\det\left(\bI + \frac{P}{p}\bHr\bHe^\sharp\bHr^\dagger\right) + o(1),
\label{eq:maxQ}
\end{align}

To obtain \eqref{eq:maxQ}, let
\begin{equation}
\g = \sigma_{\max}(\bF_{31}\bF_{31}^\dagger + \bF_{32}\bF_{32}^\dagger)
\label{eq:g-def}
\end{equation}
denote the largest singular value of the matrix
$\bF_{31}\bF_{31}^\dagger + \bF_{32}\bF_{32}^\dagger$.  Since
$\log\det(\cdot)$ is increasing on the cone of positive semidefinite
matrices, we have
\begin{align}
\max_{\substack{\bQ\succeq \bzero: \\ \tr(\bQ)\le P}}
&\log\det(\bI + \bF_{31}\bF_{31}^\dagger + \bF_{32}\bF_{32}^\dagger + \bOm_3^{-1}\bQ \bOm_3^{-\dag}) \notag\\
&\le\max_{\substack{\bQ\succeq \bzero: \\ \tr(\bQ)\le
      P}}\log\det((1+\g)\bI + \bOm_3^{-1}\bQ \bOm_3^{-\dag})
  \label{eq:increasing_Matrices}\\ 
&= \log\det\left((1+\g)\bI + \frac{P}{p} \bOm_3^{-1}
  \bOm_3^{-\dag}\right) + o(1) \label{eq:Waterfilling_Gains}\\ 
&= \log\det\left(\bI + \frac{P}{p} \bOm_3^{-1}
  \bOm_3^{-\dag}\right) + o(1)\notag\\ 
&= \log\det\left(\bI + \frac{P}{p}\bHr
  \bHe^\sharp\bHr^\dagger\right) + o(1),
\label{eq:HrHe_reln} 
\end{align}
where \eqref{eq:increasing_Matrices} follows from the fact that
$\g\bI -\bF_{31}\bF_{31}^\dagger - \bF_{32}\bF_{32}^\dagger \succeq \bzero$,
and \eqref{eq:Waterfilling_Gains} follows from the fact that
water-filling provides a vanishingly small gain over flat power
allocation when the channel matrix has a full rank (see, e.g.,
\cite{martinian:04}), and \eqref{eq:HrHe_reln} follows from
\eqref{eq:HrHe_perp}.

\subsection{Analysis of the Masked MIMO Transmission Scheme}
\label{sec:masked-mimo-anal}

To establish \eqref{eq:Rsn}, we focus on the two terms in the argument
of \eqref{eq:CK}, obtaining
\begin{equation} 
I(\rvbu;\rvbyr) = \log\det(\bI + \Pt\bHr\bHr^\dagger)
=\log\det(\bI + \Pt\bHrsv^{2}),
\label{eq:Iuyr}
\end{equation}
where we have used \eqref{eq:bHr-svd} to obtain the second equality,
and
\begin{equation} 
I(\rvbu;\rvbye) = h(\rvbye) - h(\rvbye|\rvbu)
\label{eq:Iuye}
\end{equation}
with
\begin{equation} 
h(\rvbye) = \log\det(\bI + \Pt\bHe\bHe^\dagger)
\label{eq:hye}
\end{equation}
and
\begin{align} 
h(\rvbye|\rvbu) 
&= \log\det(\bI + \Pt\bHe \bVn\bVn^\dagger \bHe^\dagger)\notag\\
&= \log\det(\bI + \Pt\bHe (\bI -\bVr\bVr^\dagger)
\bHe^\dagger)\label{eq:use-mercer}\\
&= \log\det(\bI + \Pt (\bI - \bVr\bVr^\dagger) \bHe^\dagger\bHe),
\label{eq:hyeu}
\end{align}
where to obtain \eqref{eq:use-mercer} we have used that
$\bVr\bVr^\dagger+\bVn\bVn^\dagger=\bI$ since $\begin{bmatrix} \bVr &
\bVn \end{bmatrix}$ is unitary, and where to obtain \eqref{eq:hyeu} we
have used that $\det(\bI +\bA \bB) = \det(\bI + \bB\bA)$ for any $\bA$
and $\bB$ of compatible dimensions.

In turn, substituting \eqref{eq:hye} and \eqref{eq:hyeu} into \eqref{eq:Iuye}
we obtain, with some algebra,
\begin{align} 
I(\rvbu;\rvbye) 
&=-\log\det(\bI - \Pt(\bI
+\Pt\bHe^\dagger\bHe)^{-1}(\bVr\bVr^\dagger\bHe^\dagger\bHe)) \notag\\
&=-\log\det(\bI - \Pt \bVr^\dagger\bHe^\dagger\bHe(\bI +
\Pt\bHe^\dagger\bHe)^{-1} \bVr) ) \notag\\ 
&=-\log\det(\bVr^\dagger(\bI + \Pt\bHe^\dagger\bHe)^{-1}\bVr).
\label{eq:Iuye-exp}
\end{align}
Finally, using \eqref{eq:Iuyr} and \eqref{eq:Iuye-exp} in the argument
of \eqref{eq:CK}, and again using \eqref{eq:bHr-svd}, we obtain
[cf.~\eqref{eq:Rsn}]
\begin{align*} 
&R_\mrm{SN}(P)\notag\\
&\ = \log\det(\bI + \Pt\bHrsv^{2})  + \log\det(\bVr^\dagger(\bI +
  \Pt\bHe^\dagger\bHe)^{-1}\bVr) \notag\\ 
&\ =\log\det(\Pt\bI + \bHrsv^{-2}) \notag\\
&\ \qquad\quad +\log\det(\bU\bHrsv\bVr^\dagger(\bI + \Pt
  \bHe^\dagger\bHe)^{-1}\bVr\bHrsv\bU^\dagger) \notag\\ 
&\ =\log\det(\Pt\bI + \bHrsv^{-2})+\log\det(\bHr(\bI + \Pt\bHe^\dagger\bHe)^{-1}\bHr^\dagger),
\end{align*}
as required.

Finally, to establish the first equality in \eqref{eq:Rsn_highSNR}, we
take the limit $\Pt\rightarrow\infty$ in \eqref{eq:Rsn}.  In
particular, we have
\begin{align}
&R_\mrm{SN}(P) \notag\\
&\quad= \log\det(\bI + \Pt^{-1}\bHrsv^{-2}) \notag\\
&\quad\qquad {} +\log\det(\bHr(\Pt^{-1}\bI +
  \bHe^\dagger\bHe)^{-1}\bHr^\dagger), \notag\\ 
&\quad= O(\Pt^{-1}) 
+ \log\det(\bHr((\bHe^\dagger\bHe)^{-1} 
+ O(\Pt^{-1}))\bHr^\dagger) \label{eq:fact1} \\ 
&\quad= \log\det(\bHr(\bHe^\dagger\bHe)^{-1}\bHr^\dagger)\notag\\ 
&\quad\qquad {} + \log\det(\bI +
(\bHe^\dagger\bHe)^{-1/2} O(\Pt^{-1})(\bHe^\dagger\bHe)^{-\dag/2})\notag\\ 
&\quad= \log\det(\bHr(\bHe^\dagger\bHe)^{-1}\bHr^\dagger) + O(\Pt^{-1}),
\label{eq:fact2}
\end{align}
where to obtain \eqref{eq:fact1} we have used that $(\eps\bI +
\bM)^{-1} = \bM^{-1} + O(\eps)$ as $\eps\rightarrow0$ for any
invertible $\bM$ \cite{petersenPedersen}, and where we 
have also used that $\log\det(\bI + \bW)$ is continuous
in the entries of $\bW$.

\section{MIMOME Channel Scaling Laws}
\label{sec:Scaling}

We first verify Claim~\ref{lem:zeroCapCond}, then use it to establish
Corollary~\ref{corol:AsympZeroCondn}.

\begin{IEEEproof}[Proof of Claim~\ref{lem:zeroCapCond}]
Clearly, $\sigma_{\max}(\bHr,\bHe) = \infty$ when
[cf.~\eqref{eq:cSr-def}] $\cSr \neq \varnothing$.  Otherwise, it is
known (see, e.g., \cite{golubVanLoan}) that $\sigma_{\max}(\cdot)$ is
the largest generalized singular value of $(\bHr,\bHe)$ as defined in
\eqref{eq:GSV-defn}.

To establish that the secrecy capacity is zero whenever
$\sigma_{\max}(\bHr,\bHe)\le 1$, it suffices to consider the high-SNR
secrecy capacity \eqref{eq:highSNRCapGen} when $\bHe$ has full
column-rank, which is clearly zero whenever $\sigma_{\max}\le 1$.

When $\sigma_{\max}(\bHr,\bHe) > 1$, there exists a vector $\bv$ such
that $\|\bHr\bv\| > \|\bHe\bv\|$.  Then, choosing $\rvbx = \rvbu \sim
\CN(\bzero, P\bv\bv^\dagger)$ in the argument of \eqref{eq:CK} yields
a strictly positive rate $R(P)$, so $C(P) \ge R(P) > 0$ for all $P>
0$.
\end{IEEEproof}

Combining Claim~\ref{lem:zeroCapCond} and
Fact~\ref{fact:LargestGSVConv} below, which is established in
\cite[p.~642]{silverStein85}, yields
Corollary~\ref{corol:AsympZeroCondn}.

\begin{fact}[\cite{silverStein85,BaiSilverstein:95}]
Suppose that $\bHr$ and $\bHe$ have i.i.d.\ $\CN(0,1)$ entries.  Let
$\Nr,\Ne,\Nt\rightarrow \infty$, while keeping $\Nr/\Ne=\g$ and
$\Nt/\Ne=\bt$ fixed.  Then if $\bt < 1$,
\begin{equation}
\sigma_{\max}(\rvbHr,\rvbHe)
\asconv \g\left[
\frac{1 + \sqrt{1-(1-\bt)\left(1-\frac{\bt} \g\right)}}{1-\bt}
\right]^2.
\label{eq:f1eq}
\end{equation}
\label{fact:LargestGSVConv}
\end{fact}

\section{Concluding Remarks}
\label{sec:conclusions}

This paper resolve several open questions regarding secure
transmission with multiple antennas.  First, it establishes the
existence of a computable expression for the secrecy capacity of the
MIMOME channel.  Second, it establishes that a Gaussian input
distribution optimizes the secrecy capacity expression of Csisz{\'a}r
and K{\"o}rner for the MIMOME channel, and thus that capacity is
achieved by Gaussian wiretap codes.  Third, it establishes the optimum
covariance structure for the input, exploiting hidden convexity in the
problem.  Nevertheless, many questions remain that are worth
exploring.  As one example, it remains to be determined whether such
developments based on Sato's bounding techniques be extended beyond
sum-power constraints, as the channel enhancement based approach of
\cite{LiuShamai:07} can.

In addition, our analysis highlights the useful role that the GSVD
plays both in calculating the capacity of the MIMOME channel in the
high-SNR regime, and in designing codes for approaching this capacity.
At the same time, we observed that a simple, semi-blind masked MIMO
scheme can be arbtrarily far from capacity.  However, for the special
case of the MISOME channel, \cite{khistiWornell:07} shows that the
corresponding masked beamforming scheme achieve rates close to
capacity at high SNR.  Thus, it remains to be determined whether there
are better and/or more natural generalizations of the masked
beamforming scheme for the general MIMOME channel.  This warrants
further investigation.

More generally, semi-blind schemes have the property that they require
only partial knowledge of the channel to the eavesdropper.  Much
remains to be explored about what secrecy rates are achievable with
such partial information.  One recent work in this area~\cite{khisti:10} illustrates the use of interference alignment techniques for the compound extension of the multi-antenna wiretap channel. Another recent work~\cite{chandar2010}, studies a constant-capacity compound
wiretap channel model  which again captures the constraint
that the transmitter only knows the capacity (or an upper bound on the
capacity) of the channel to the eavesdropper.  Further insights may arise
from considering other multiple eavesdropper scenarios with limited or
no collusion.

Finally, we characterize when an eaversdropper can prevent secure
communication, i.e., drive the secrecy capacity to zero.  Our scaling
laws on antenna requirements and their optimal distribution in limit
of many antennas provide convenient rules of thumb for system
designers, as the results become independent of the channel matrices
in this limit.  However, it remains to quantify for what numbers of
antennas these asymptotic results become meaningful predictors of
system behavior.   As such, this represents yet another useful direction
for further research.

\section*{Acknowledgement}

We thank Ami Wiesel for interesting discussions and help with
numerical optimization of the saddle point expression in
Theorem~\ref{thm1}.

\appendices

\section{Proof of Claim~\ref{claim:noiseSing}}
\label{app:equiv-obs}

To begin, 
\begin{align}
I(\rvbx;\rvbyr|\rvbye)
&=I(\rvbx; \bU_1^\dagger\rvbyr,
\bU_2^\dagger\rvbyr|\rvbye) \label{eq:use-unitary}\\ 
&=I(\rvbx; \rvbytr,
\bU_1^\dagger\rvbyr-\bV_1^\dagger\rvbye|\rvbye) \notag \\ 
&=I(\rvbx; \rvbytr,\bT\rvbx|\rvbye), \label{eq:Ixyinf}
\end{align}
where \eqref{eq:use-unitary} follows from the fact that
$\begin{bmatrix} \bU_1 & \bU_2\end{bmatrix}$ is unitary, and where
\eqref{eq:Ixyinf} follows from substituting for $\rvbyr$ and $\rvbye$
from \eqref{eq:channel}, using \eqref{eq:bT-def}, from and the fact that
\begin{equation} 
\bU_1^\dagger \rvbzr \aseq \bV_1^\dagger\rvbze,
\label{eq:noiseSing1a}
\end{equation}
since
\begin{equation*} 
\cov(\bU_1^\dagger\rvbzr,\bV_1^\dagger\rvbze) = 
E[\bU_1^\dagger\rvbzr\rvbze^\dagger\bV_1]= \bU_1^\dagger\bPh\bV_1=
\bI.
\end{equation*}
Now when $I(\rvbx;\rvbyr|\rvbye) < \infty$, we have from
\eqref{eq:Ixyinf} that $\bT\rvbx = \bzero$, so
$I(\rvbx;\rvbyr|\rvbye)=I(\rvbx;\rvbytr|\rvbye)$, establishing
\eqref{eq:noiseSing3a}.

Similarly,
\begin{align}
I(\rvbx;\rvbye|\rvbyr)
&=I(\rvbx; \bV_1^\dagger\rvbye,
\bV_2^\dagger\rvbye|\rvbyr) \label{eq:use-unitary-again}\\ 
&=I(\rvbx; \rvbyte,
\bV_1^\dagger\rvbye-\bU_1^\dagger\rvbyr|\rvbyr) \notag \\ 
&=I(\rvbx; \rvbyte,\bT\rvbx|\rvbyr), \label{eq:Ixyinf-var}
\end{align}
where we have used that $\begin{bmatrix} \bV_1 & \bV_2\end{bmatrix}$
is unitary to obtain \eqref{eq:use-unitary-again} and
\eqref{eq:noiseSing1a} to obtain \eqref{eq:Ixyinf-var}.
When $I(\rvbx;\rvbye|\rvbyr) < \infty$, we have from
\eqref{eq:Ixyinf-var} that $\bT\rvbx = \bzero$, so
$I(\rvbx;\rvbye|\rvbyr)=I(\rvbx;\rvbyte|\rvbyr)$, establishing
\eqref{eq:noiseSing3b}.

To verify the ``only if'' statement of the last part of the claim, 
when $I(\rvbx;\rvbyr|\rvbye) = \infty$, we 
expand \eqref{eq:Ixyinf} via the chain rule to obtain
\begin{equation} 
I(\rvbx;\rvbyr|\rvbye) 
=I(\rvbx; \rvbytr|\rvbye) + I(\rvbx; \bT\rvbx|\rvbytr,\rvbye),
\label{eq:I-inf-exp}
\end{equation}
and note that if $\bT\rvbx\aseq\bzero$ then the second term on the
right-hand side of \eqref{eq:I-inf-exp} is zero.  But the first term
on the right-hand side is finite, so $\cov(\bT\rvbx)\neq\bzero$, i.e.,
\eqref{eq:inf-I-cond}, holds.

To verify the ``if'' statement of the last part of the claim, we use
the chain rule to write
\begin{align}
I(\rvbx;\rvbyr|\rvbye) 
&= I(\rvbx;\rvbytr,\bT\rvbx|\rvbye) \notag\\
&\ge I(\rvbx;\bT\rvbx|\rvbye) \notag\\
&\ge I(\rvbx;\bT\rvbx)-I(\rvbx;\rvbye), \label{eq:lastline}
\end{align}
and note that the first term in \eqref{eq:lastline} is infinite when
$\cov(\bT\rvbx)\neq\bzero$, while the second term is finite.
\hfill\IEEEQEDclosed

\section{Optimizing $\Rp(p_\rvbx,\bK_\bPh)$ Over $p_\rvbx$ with
  Singular $\bK_\bPh$} 
\label{GaussOpt}

To establish that $I(\rvbx;\rvbyr|\rvbye)$ with $\rvbz \sim
\CN(\bzero, \bK_\bPh)$ for singular $\bK_\bPh$ is maximized subject to
the constraint $\cov(\rvbx)=\bK_P$ when $\rvbx$ is Gaussian (hence,
justifying \eqref{eq:GaussMax} in this case), we exploit
Claim~\ref{claim:noiseSing}.

In particular, if for all $p_\rvbx$ meeting the covariance constraint
we have $I(\rvbx;\rvbyr|\rvbye)<\infty$, then we can use
\eqref{eq:noiseSing3a}, expanding and bounding $I(\rvbx;\rvbytr|
\rvbye)$ in the same manner as
\eqref{eq:mutInfExp}--\eqref{eq:mutInfExp2}, with $\rvbytr$,
$\rvbztr$, $\bLat\defeq\bU_2^\dagger\bLa\bU_2$ (the error covariance
in the MMSE estimate of $\rvbytr$ from $\rvbye$), and
$\tbPhr=\bU_2^\dagger\bPh$ [cf.\ \eqref{eq:equiv-noise-cov}] replacing
$\rvbyr$, $\rvbzr$, $\bLa$, and $\bPh$, respectively.  Specifically,
we obtain that
\begin{equation} 
I(\rvbx;\rvbytr | \rvbye)  = h(\rvbytr | \rvbye) - h(\rvbztr | \rvbze),
\label{eq:It-exp}
\end{equation}
is maximized when $\rvbx$ is Gaussian.  

If, instead, there exists a $p_\rvbx$ satisfying the covariance
constraint such that $I(\rvbx;\rvbyr|\rvbye)=\infty$, then by the
``only if'' part of the last statement of Claim~\ref{claim:noiseSing}
we have that \eqref{eq:inf-I-cond} holds.  But by the ``if'' part of
the same statement we know that $I(\rvbx;\rvbyr|\rvbye)=\infty$ for
any $p_\rvbx$ such that \eqref{eq:inf-I-cond} holds, and in particular
we may choose $p_\rvbx$ to be Gaussian.  \hfill\IEEEQEDclosed

\iffalse
then using
the entropy maximizing property of Gaussian distributions with
\eqref{eq:It-exp} implies that the choice $p_\rvbx= \CN(\bzero,\bK_P)$
also results in $I(\rvbx;\rvbytr|\rvbye)=\infty$.  And since for any
$p_\rvbx$ we have $I(\rvbx;\rvbyr|\rvbye)\ge I(\rvbx;\rvbytr|\rvbye)$,
it follows that $I(\rvbx;\rvbyr|\rvbye)=\infty$ for the Gaussian
choice.

Finally, if $\bK_P$ is such that \eqref{eq:inf-I-cond} holds, then by
the last statement of Claim~\ref{claim:noiseSing},
$I(\rvbx;\rvbyr|\rvbye)=\infty$ regardless of the whether $p_\rvbx$ is
Gaussian or not, so all distributions are maximizing, including the
choice $p_\rvbx=\CN(\bzero,\bK_P)$.  \hfill\IEEEQEDclosed
\fi

\section{Proof of Lemma~\ref{lem:saddlePointProperty-a} for Singular
  $\obK_\bPh$} 
\label{app:KphCondnSing}

We begin with the following:
\begin{claim}
\label{claim:Kph-svd}
There exists a matrix $\bHtt$ such that the combined channel matrix
\eqref{eq:bHt-def} can be expressed in the form
\begin{equation}
\label{eq:HGWrel}
\bHt = \bW \bHtt,
\end{equation}
where
\begin{equation}
\obK_\bPh= \bW \obXi\bW^\dagger,
\label{eq:Wdecomp}
\end{equation}
is the compact singular value decomposition of $\obK_\bPh$, i.e., 
where $\bW$ has orthogonal columns ($\bW^\dagger\bW = \bI$), and
the diagonal matrix $\obXi$ has strictly positive diagonal
entries.
\end{claim}
Hence, the column space of $\bHt$ is a subspace of the column space of
$\bW$.

\begin{IEEEproof}
We establish our result by contradiction.  Suppose the claim were
false.  Then clearly $I(\rvbx;\rvbyr,\rvbye)=\infty$ when we choose
$\rvbx=\rvt \bups$ where $\bups\in\Null(\bW)$ and $\var\rvt>0$, which
implies that
\begin{equation*} 
\Rp(\bK_P,\obK_\bPh) = I(\rvbx;\rvbyr|\rvbye) = I(\rvbx;\rvbyr,\rvbye)
- I(\rvbx;\rvbye) = \infty,
\end{equation*}
since $I(\rvbx;\rvbye)<\infty$ as $\cov(\rvbze)=\bI$ is
nonsingular.   Hence,
\begin{equation}
\Rp(\obK_P,\obK_\bPh) = \max_{\bK_P\in\cK_P}\Rp(\bK_P,\obK_\bPh) =
\infty.
\label{eq:Rp-inf}
\end{equation}

But from \eqref{eq:finite-saddle} in
Lemma~\ref{lem:saddlePointExistence} we know $\Rp(\obK_P,\obK_\bPh)
<\infty$, which contradicts \eqref{eq:Rp-inf} and hence
\eqref{eq:HGWrel} must hold.
\end{IEEEproof}

Using Claim~\ref{claim:Kph-svd}, we see that in this case the original
channel \eqref{eq:channel} with $\cov(\rvbz)=\bK_\bPh$ can be replaced
with the equivalent combined channel
\begin{equation} 
\rvbyt = \bHtt \rvbx + \rvbzt
\end{equation}
where
\begin{equation*} 
\rvbyt \defeq \bW^\dagger \begin{bmatrix} \rvbyr \\ \rvbye 
\end{bmatrix},\qquad \rvbzt \defeq \bW^\dagger
\rvbz,
\end{equation*}
with $\cov(\rvbz)=\bXi$.  Hence, we can write
\begin{equation*} 
\Rp(\obK_P,\obK_\bPh) = I(\rvbx;\rvbyr,\rvbye) - I(\rvbx;\rvbye),
\end{equation*}
where
\begin{equation} 
I(\rvbx;\rvbyr,\rvbye) = I(\rvbx;\rvbyt) =
\log\frac{\det(\bXi + \bHtt\obK_P\bHtt^\dagger)}{\det(\bXi)}
\end{equation}
and
\begin{equation} 
I(\rvbx;\rvbye) = \log\det(\bI +\bHe\obK_P\bHe^\dagger).
\end{equation}
But from the saddle point property it follows that $\obXi$
can be expressed as
\begin{equation}
\obXi = \argmin_{\{\bXi \,:\,\bW \bXi \bW^\dagger \in \bK_\bPh \}}
\log\frac{\det(\bXi+\bHtt\obK_P\bHtt^\dagger)}{\det(\bXi)}.
\label{eq:noiseSingularMinProblem}
\end{equation}

\begin{comment}
\begin{align}R_\Omega(\bXi)&= I(\rvbx; \bG\rvbx + \rvbz_\Omega)\notag\\
&=I(\rvbx; \bW\bG\rvbx + \bW\rvbz_\Omega)\label{eq:Wmult}\\ &=I(\rvbx;
\bHt\rvbx + \rvbz) \label{eq:omNoiseProp}\\ &=I(\rvbx;
\rvbyr,\rvbye)\notag\\ &= \Rp(\obK_P,\bW\Omega\bW^\dagger) +
\log\det(\bI + \bHe\obK_P\bHe^\dagger)\label{eq:R+RomegRel}\end{align}
where \eqref{eq:Wmult} follows from the fact that $\bW$ has a full
column-rank, so right multiplication by $\bW$ preserves the mutual
information and \eqref{eq:omNoiseProp} follows from \eqref{eq:HGWrel}
and the fact that $\rvbz_\Omega \in \cK_\bXi$.  If there exists any $\Wh$
such that $R_\Omega(\Wh) <R_\Omega(\obXi)$ then from
\eqref{eq:R+RomegRel} we have that 
\begin{equation*} 
\Rp(\obK_P,\obK_\bPh) > \Rp(\obK_P,\bW\Wh\bW^\dagger)
\end{equation*}
which [cf.~\eqref{eq:RKbKphSaddle}] contradicts that $\obK_\bPh$ is a
saddle point solution.
\end{comment}

In turn, the KKT conditions associated with the optimization
\eqref{eq:noiseSingularMinProblem} are
\begin{equation*}
\obXi^{-1}-(\obXi+\bHtt\obK_P\bHtt^\dagger)^{-1}=\bW^\dagger\bUp\bW,
\end{equation*}
or, equivalently,
\begin{equation} 
\bHtt\obK_P\bHtt^\dagger 
= \obXi\bW^\dagger \bUp \bW(\obXi + \bHtt\obK_P\bHtt^\dagger),
\label{eq:OmegKKT}
\end{equation}
where the dual variable $\bUp$ is of the same block diagonal form as
in the nonsingular case, viz., \eqref{eq:UpsDefn}.  Multiplying the
left- and right-hand sides of \eqref{eq:OmegKKT} by $\bW$ and
$\bW^\dagger$, respectively, and using \eqref{eq:HGWrel} and
\eqref{eq:Wdecomp} we obtain \eqref{eq:noiseKKTCondnb}.  Thus, the
remainder of the proof uses the arguments following
\eqref{eq:noiseKKTCondnb}
%\eqref{eq:Hrr}--\eqref{eq:elim-end} 
in the proof for the nonsingular case to establish the desired result.
\hfill\IEEEQEDclosed

\section{Proof of Proposition~\ref{claim:HSaddlePoint}}
\label{app:KKTOptimalityProof}

Consider first the right-hand side of \eqref{eq:secondOpt}.  Since
$h(\rvbyr-\obTh\rvbye)$
is concave in $\bK_P \in \cK_P$ and differentiable over $\cK_P$, the
KKT conditions associated with the Lagrangian
\begin{multline} 
\cL_\bTh(\bK_P,\lambda,\bPs) \\
= h(\rvbyr-\obTh\rvbye)+\tr(\bPs\bK_P) - \lambda(\tr(\bK_P)-P) 
\end{multline}
are both necessary and sufficient, i.e., $\bK_P$ is a solution to the
right-hand side of \eqref{eq:secondOpt} if and only if there exists a
$\lambda \ge 0$ and $\bPs \succeq \bzero$ such that
\begin{equation}
\begin{gathered}
(\bHr-\obTh\bHe)^\dagger \bGa(\bK_P,\obK_P)^{-1}(\bHr-\obTh\bHe) + \bPs = \lambda\bI,\\
\tr(\bPs \bK_P)=0, \quad \text{and} \quad \lambda(\tr(\bK_P)-P)=0,\label{eq:HmaxKKT}
\end{gathered}
\end{equation}
where
\begin{align}
&\bGa(\bK_P,\obK_P) \notag\\
&\ \defeq \cov(\rvbyr - \obTh\rvbye)\quad \notag\\
&\ = \bI + \obTh\obTh^\dagger -\obTh\obPh^\dagger-\obPh\obTh^\dagger \notag\\
&\qquad {} + (\bHr-\obTh\bHe)\bK_P(\bHr-\obTh\bHe)^\dagger. 
\label{eq:GammaDef}
\end{align}

Considering next the left-hand side of \eqref{eq:secondOpt}, to which
$\obK_P$ is a solution, we have, from the associated KKT conditions,
that there exists $\lambda' \ge 0$ and $\bPs' \succeq \bzero$ such that
\begin{equation}
\begin{aligned}
\nabla_{\bK_P} h(\rvbyr - \bTh(\bK_P) \rvbye)
\bigr|_{\bK_P=\obK_P} +\bPs' 
= \lambda'\bI\\ 
\tr(\bPs' \obK_P)=0,\quad\text{and}\quad\lambda'(\tr(\obK_P)-P)=0,
\end{aligned}
\label{eq:OptKpKKT}
\end{equation}
where $\bTh(\bK_P)$ is as defined in \eqref{eq:bTh-def}.

Thus, it remains to show that \eqref{eq:HmaxKKT} and
\eqref{eq:OptKpKKT} are identical when $\bK_P=\obK_P$.
Focusing on the first equation in \eqref{eq:OptKpKKT}, we have
\begin{align}
&\nabla_{\bK_P} h(\rvbyr - \bTh(\bK_P) \rvbye) \bigr|_{\bK_P=\obK_P}\notag\\
&\ =\nabla_{\bK_P} h(\rvbyr | \rvbye) \bigr|_{\bK_P=\obK_P}\notag\\
&\ = \nabla_{\bK_P} \left\{ h(\rvbyr , \rvbye) - h(\rvbye)\right\} \bigr|_{\bK_P=\obK_P}\notag\\
&\ = \bHt^\dagger(\bHt\obK_P\bHt^\dagger + \obK_\bPh)^{-1}\bHt -
  \bHe^\dagger(\bI + \bHe\obK_P\bHe^\dagger)^{-1}\bHe.  
\label{eq:nabla-exp}
\end{align}

In turn, substituting for $\bHt$ and $\obK_\bPh$ from
\eqref{eq:bHt-def} and \eqref{eq:oPh-def}, and using
\eqref{eq:obTh-def}, the first matrix inverse in \eqref{eq:nabla-exp}
can be expressed in the form
\begin{align}
&(\bHt\obK_P\bHt^\dagger+\obK_\bPh)^{-1}\notag\\
&=\begin{bmatrix} \bI + \bHr\obK_P\bHr^\dagger & \obPh + \bHr \obK_P
  \bHe^\dagger\\ \obPh^\dagger + \bHr\obK_P\bHe^\dagger & \bI +
  \bHe\obK_P\bHe^\dagger \end{bmatrix}^{-1}\notag\\ 
&=\begin{bmatrix} \bLa(\bK_P)^{-1} & -\bLa(\bK_P)^{-1}\obTh\\
-\obTh^\dagger\bLa(\bK_P)^{-1} &
(\bI\!+\!\bHe\obK_P\bHe)^{-1}\!+\!\obTh^\dagger\bLa(\bK_P)^{-1}\obTh \end{bmatrix},  
\label{eq:use-mil}
\end{align}
where $\bLa(\bK_P)$ is as defined in \eqref{eq:obLa-def}, and where we
have used the matrix inversion lemma (see, e.g.,
\cite{petersenPedersen}).  Substituting \eqref{eq:use-mil} into
\eqref{eq:nabla-exp}, and using the notation \eqref{eq:obLa-def},
yields, after some simplification, 
\begin{align}
&\nabla_{\bK_P} h(\rvbyr - \bTh(\bK_P) \rvbye) \bigr|_{\bK_P=\obK_P}\notag\\
&\quad=\bHt^\dagger(\obK_\bPh + \bHt\obK_P\bHt^\dagger)^{-1}\bHt -
  \bHe^\dagger(\bI + \bHe\obK_P\bHe^\dagger)^{-1}\bHe  \notag\\
&\quad= (\bHr-\obTh\bHe)^{\dagger} \obLa^{-1} (\bHr-\obTh\bHe).
\label{eq:Rmmse}
\end{align}

Comparing \eqref{eq:Rmmse} with the first equation in
\eqref{eq:HmaxKKT}, we see that it remains only to show that
$\bGa(\obK_P,\obK_P)=\obLa$, which is verified as follows.  First,
$\obTh\rvbye$ is the MMSE estimate of $\rvbyr$ from $\rvbye$ when
$\bK_P=\obK_P$, and $\bGa(\obK_P,\obK_P)=\cov(\rvbyr - \obTh\rvbye) =
\cov(\rvbyr|\rvbye)$ is the error covariance associated with the
estimate.  But by definition [cf.~\eqref{eq:obLa-def}]
$\obLa=\cov(\rvbyr|\rvbye)$ is also the error covariance associated
with the MMSE estimate when $\bK_P=\obK_P$, so the conclusion follows.
\hfill\IEEEQEDclosed

\section{Proof of Lemma~\ref{lem:saddlePointProperty-b} for Singular
  $\obK_\bPh$} 
\label{app:KphSingRankCondn}

First, note that via \eqref{eq:saddlePointProp} with
\eqref{eq:finite-saddle}, we have that $\Rp(\bK_P,\obK_\bPh)
=I(\rvbx;\rvbyr|\rvbye) < \infty$ for all $\bK_P\in\cK_P$.  Hence, via
\eqref{eq:noiseSing3a} of Claim~\ref{claim:noiseSing} we have
\begin{equation}
\Rp(\bK_P,\obK_\bPh) = I(\rvbx;\rvbytr| \rvbye),\quad
\forall\,\bK_P\in\cK_P,
\label{eq:noiseSing3}
\end{equation}
with the equivalent observations $\rvbytr$ as given by
\eqref{eq:rvbytr-def} with \eqref{eq:bHtr-rvbztr-def}.  Moreover, the
noise cross-covariance $\tbPhr =\bU_2^\dagger\obPh$ [cf.\
\eqref{eq:equiv-noise-cov}] in the equivalent channel model has all
its singular values strictly less than unity, i.e., the associated
$\bK_\tbPhr$ is nonsingular.

Thus, we can apply to this equivalent model the arguments of the proof
of Lemma~\ref{lem:saddlePointProperty-b} for the nonsingular case.  In
particular, from \eqref{eq:use-saddle-first} onwards we replace
$\rvbyr$ with $\rvbytr$, we replace
$\bTh(\bK_P)$ and $\obTh$ with, respectively,
[cf. \eqref{eq:bTh-def},\eqref{eq:obTh-def}]
\begin{equation}
\tbTh(\bK_P) \defeq
(\bHtr\bK_P\bHe^\dagger + \tbPhr)(\bI + \bHe\bK_P\bHe^\dagger)^{-1} = 
\bU_2^\dagger \bTh(\bK_P)
\label{eq:tbTh-def}
\end{equation}
and
\begin{equation} 
\tobTh \defeq \tbTh(\obK_P) = \bU_2^\dagger \obTh,
\label{eq:tobTh-def}
\end{equation}
which is the coefficient in the MMSE estimate of $\rvbytr$ from $\rvbye$,
and we replace $\obHeff$ and $\obJ$ with, respectively, [cf.\
\eqref{eq:bHeff-def}]
\begin{equation*}
\tbHeff \defeq \tbJ^{-1/2}(\bHtr-\tobTh\bHe)
\end{equation*}
and
\begin{equation*} 
\tbJ 
\defeq (\bI-\tbPh\tbPh^\dagger) +
(\tobTh-\tbPhr)(\tobTh-\tbPhr)^\dagger
= \bU_2^\dagger \obJ \bU_2,
\end{equation*}
noting that $\tbJ\succ\bzero$ since $\bK_\tbPhr\succ\bzero$.   With
these changes, and with the SVD
\begin{equation*}
\tbHeff = \tbA \tbSieff \tbB^\dagger
%\label{eq:tHeff-svd}
\end{equation*}
replacing \eqref{eq:Heff-svd}, the arguments apply and it follows that $(\bHtr-\tobTh \bHe)\bS$ has a full column rank.  Since $$(\bHtr-\tobTh \bHe)\bS=\bU_2^\dagger(\bHr - \obTh\bHe)\bS$$ it then follows that $(\bHr-\bTh\bHe)\bS$ has a full column rank. 
\hfill\IEEEQEDclosed

\section{Proof of Lemma~\ref{lem:upperBoundSimplified} for Singular
  $\obK_\bPh$} 
\label{app:upper_Bound_Simplified_Singular}

Consider first the case in which $\bHr \neq \obTh\bHe$, and
note that
\begin{equation*} 
\Rp(\obK_P,\obK_\bPh) - \Rm(\obK_P) = I(\rvbx;\rvbye|\rvbyr) < \infty,
\end{equation*}
where the equality is reproduced from \eqref{eq:I-gap}, and where the
inequality follows from \eqref{eq:finite-saddle} and that
$\Rm(\obK_P)\ge0$.  Hence, applying \eqref{eq:noiseSing3b} from
Claim~\ref{claim:noiseSing}, we have
\begin{equation*}
\Rp(\obK_P,\obK_\bPh) - \Rm(\obK_P) = I(\rvbx;\rvbyte|\rvbyr),
\end{equation*}
with the equivalent observations $\rvbyte$ as given by
\eqref{eq:rvbyte-def} with \eqref{eq:bHte-rvbzte-def}.  Moreover, the
noise cross-covariance 
\begin{equation} 
\tbPhe \defeq E[\rvbzr\rvbzte^\dagger] = \obPh \bV_2
\label{eq:tbPhe-def}
\end{equation}
in the equivalent channel model has all its singular values strictly
less than unity, i.e., the associated $\bK_\tbPhe$ is nonsingular.

Thus, we can apply to this equivalent model the corresponding
arguments of the proof of Lemma~\ref{lem:upperBoundSimplified} for the
nonsingular case.  In particular, from \eqref{eq:I-gap} onwards we
replace $\rvbye$ and $\rvbze$ with, respectively, $\rvbyte$ and
$\rvbzte$, we replace $\bLab$ with
\begin{align*} 
&\tbLab = \notag\\
&\ \ \bI+\bHte\obK_P\bHte^\dagger \notag\\
&\quad {} - (\tbPhe^\dagger + \bHte\obK_P\bHr^\dagger) 
(\bI + \bHr\obK_P\bHr^\dagger)^{-1} (\tbPhe+\bHr\obK_P\bHte^\dagger))
  \notag\\ 
&\quad\quad  = \bV_2^\dagger \bLab \bV_2,
\end{align*}
which is the backward error covariance associated with the linear MMSE
estimate of $\rvbyte$ from $\rvbyr$, and we replace the use of
\eqref{eq:degradedness} in \eqref{eq:use-nc-1} and \eqref{eq:use-nc-2}
with its form for the equivalent channel, viz., for all full
column-rank $\obS$ such that $\obS\obS^\dagger=\obK_P$,
\begin{equation*} 
\tbPhe^\dagger \bHr \obS 
= \bV_2^\dagger \obPh^\dagger \bHr \obS 
= \bV_2^\dagger \bHe\obS 
 = \bHte \obS,
\end{equation*}
where to obtain the first equality we have used \eqref{eq:tbPhe-def},
where to obtain the second equality we have used
Property~\ref{prop:degradedness}, and where to obtain the third
equality we have used \eqref{eq:bHte-rvbzte-def}.

Finally, consider the case in which $\bHr = \obTh \bHe$.  Since
\eqref{eq:finite-saddle} holds, so does \eqref{eq:noiseSing3a} of
Claim~\ref{claim:noiseSing}, and thus 
\begin{equation}
\Rp(\obK_P,\obK_\bPh) = I(\rvbx;\rvbytr| \rvbye),
\label{eq:noiseSing3-saddle}
\end{equation}
with the equivalent observations $\rvbytr$ as given by
\eqref{eq:rvbytr-def} with \eqref{eq:bHtr-rvbztr-def}.

Thus, we can apply to this equivalent model the corresponding
arguments of the proof of Lemma~\ref{lem:upperBoundSimplified} for the
nonsingular case.  In particular (and as in
Appendix~\ref{app:KphSingRankCondn}), from \eqref{eq:HrHecond0}
onwards \eqref{eq:noiseSing3-saddle} implies we replace $\rvbyr$ and
$\rvbzr$ with, respectively, $\rvbytr$ and $\rvbztr$, we replace
$\obPh$ with [cf.\ \eqref{eq:equiv-noise-cov}] $\tbPhr =\bU_2^\dagger
\obPh$, the coefficient in the MMSE estimate of $\rvbztr$ from
$\rvbze$, and we replace $\obTh$ with [cf.\
\eqref{eq:tbTh-def},\eqref{eq:tobTh-def}]
\begin{equation}
\tobTh  = (\bHtr\obK_P\bHe^\dagger + \tbPhr)
(\bI + \bHe\obK_P\bHe^\dagger)^{-1} = \bU_2^\dagger \obTh,
\label{eq:tobTh-repeat-def}
\end{equation}
the coefficient in the MMSE estimate of $\rvbytr$ from $\rvbye$.

Note that in obtaining the counterpart of \eqref{eq:HrHecond} we use
that $\rvbytr - \tobTh\rvbye = \rvbztr - \tobTh\rvbze$ since
\begin{equation} 
\bHtr = \bU_2^\dagger \bHr = \bU_2^\dagger \obTh \bHe = \tobTh \bHe,
\label{eq:case-equiv}
\end{equation}
where the first equality follows from \eqref{eq:bHtr-rvbztr-def}, the
second equality follows from the assumption $\bHr = \obTh \bHe$, and
the third equality from \eqref{eq:tobTh-repeat-def}.  Moreover, in
obtaining the counterpart of \eqref{eq:th-ph} we use that
$\tobTh=\tbPh$ when \eqref{eq:case-equiv} holds.  \hfill\IEEEQEDclosed

% \IEEEtriggeratref{99}
\bibliographystyle{IEEEtranS}
%\bibliography{sm}

\end{document}

%% file: gww_defs.tex
\usepackage{verbatim} % get comment environment, + new verbatim
 \usepackage{amsxtra}  % get, eg, \accentedsymbol
\DeclareMathOperator*{\argmax}{arg\,max}
\DeclareMathOperator*{\argmin}{arg\,min}

\DeclareMathOperator{\diag}{diag}
\DeclareMathOperator{\rank}{rank}
\DeclareMathOperator{\spn}{span}

\DeclareMathOperator{\tr}{tr}

\DeclareMathOperator{\var}{var}
\DeclareMathOperator{\cov}{cov}

\DeclareMathOperator{\Null}{Null}

\newcounter{actr}
{\begin{list}{(\alph{actr})}{\usecounter{actr}}}{\end{list}}

\newcounter{ictr}
{\begin{list}{(\roman{ictr})}{\usecounter{ictr}}}{\end{list}}

\newtheorem{thm}{Theorem}
\newtheorem{lemma}{Lemma}
\newtheorem{claim}{Claim}
\newtheorem{corol}{Corollary}
\newtheorem{prop}{Proposition}
\newtheorem{defn}{Definition}
\newtheorem{fact}{Fact}
\newtheorem{property}{Property}
\newenvironment{new-proof}[1]
{{\em Proof  #1: }}%
{ \noindent\qed }

\newcommand{\viz}{viz.}

\newcommand{\ds}{\displaystyle}                % abbreviation
\newcommand{\qed}{\rule[0.1ex]{1.4ex}{1.6ex}}

\newcommand{\defeq}{\stackrel{\Delta}{=}}
\newcommand{\asconv}{\stackrel{\mathrm{a.s.}}{\longrightarrow}}
\newcommand{\aseq}{\stackrel{\mathrm{a.s.}}{=}}

\newcommand{\compls}{\mathbb{C}}

\newcommand{\mrm}{\mathrm}

\newcommand{\T}{{\mathrm{T}}}

%% file: gww_chars.tex
\newcommand{\ba}{{\mathbf{a}}}

\newcommand{\bA}{{\mathbf{A}}}
\newcommand{\tbA}{{\tilde{\bA}}}

\newcommand{\bB}{{\mathbf{B}}}
\newcommand{\tbB}{{\tilde{\bB}}}

\newcommand{\bD}{{\mathbf{D}}}

\newcommand{\bF}{{\mathbf{F}}}

\newcommand{\obF}{{\bar{\bF}}}

\newcommand{\bG}{{\mathbf{G}}}

\newcommand{\bH}{{\mathbf{H}}}
\newcommand{\bHh}{{\hat{\bH}}}
\newcommand{\obH}{{\bar{\bH}}}

\newcommand{\bI}{{\mathbf{I}}}

\newcommand{\bJ}{{\mathbf{J}}}
\newcommand{\obJ}{{\bar{\bJ}}}
\newcommand{\tbJ}{{\tilde{\bJ}}}

\newcommand{\cK}{{\mathcal{K}}}
\newcommand{\bK}{{\mathbf{K}}}

\newcommand{\obK}{\bar{\bK}}

\newcommand{\cL}{{\mathcal{L}}}

\newcommand{\bM}{{\mathbf{M}}}

\newcommand{\CN}{{\mathcal{CN}}}
\newcommand{\Pt}{{\tilde{P}}}

\newcommand{\bQ}{{\mathbf{Q}}}

\newcommand{\bS}{{\mathbf{S}}}
\newcommand{\obS}{\bar{\mathbf{S}}}
\newcommand{\cS}{{\mathcal{S}}}

\newcommand{\bT}{{\mathbf{T}}}

\newcommand{\bU}{{\mathbf{U}}}

\newcommand{\bv}{{\mathbf{v}}}
\newcommand{\bV}{{\mathbf{V}}}

\newcommand{\bW}{{\mathbf{W}}}

\newcommand{\al}{\alpha}

\newcommand{\bt}{\beta}

\newcommand{\g}{\gamma}

\newcommand{\bGa}{{\mathbf{\Gamma}}}

\newcommand{\De}{\Delta}

\newcommand{\bDe}{{\mathbf{\De}}}

\newcommand{\eps}{\epsilon}

\newcommand{\bPh}{{\mathbf{\Phi}}}
\newcommand{\tbPh}{{\tilde{\bPh}}}
\newcommand{\cbPh}{{\breve{\bPh}}}

\newcommand{\obPh}{{\bar{\bPh}}}

\newcommand{\bps}{{\boldsymbol{\psi}}}
\newcommand{\bPs}{{\mathbf{\Psi}}}

\newcommand{\bTh}{{\mathbf{\Theta}}}

\newcommand{\obTh}{{\bar{\bTh}}}
\newcommand{\tbTh}{{\tilde{\bTh}}}

\newcommand{\La}{\Lambda}
\newcommand{\bLa}{{\mathbf{\La}}}
\newcommand{\obLa}{\bar{\bLa}}
\newcommand{\bLat}{\tilde{\bLa}}

\newcommand{\bXi}{{\mathbf{\Xi}}}
\newcommand{\obXi}{\bar{\bXi}}

\newcommand{\bSi}{{\mathbf{\Sigma}}}
\newcommand{\tbSi}{\tilde{\bSi}}

\newcommand{\ups}{\upsilon}
\newcommand{\bups}{{\boldsymbol{\ups}}}
\newcommand{\Ups}{\Upsilon}
\newcommand{\bUp}{{\mathbf{\Ups}}}

\newcommand{\bOm}{{\mathbf{\Omega}}}

\newcommand{\Wh}{\hat{\Omega}}

\newcommand{\bPsi}{{\mathbf{\Psi}}}
\newcommand{\bpsi}{{\boldsymbol{\psi}}}

%% file: gww_rvs.tex
\DeclareMathAlphabet{\mathbsf}{OT1}{cmss}{bx}{n}% bold sans serif
\DeclareMathAlphabet{\mathssf}{OT1}{cmss}{m}{sl}% slanted sans serif

\DeclareSymbolFont{bsfletters}{OT1}{cmss}{bx}{n}  
\DeclareSymbolFont{ssfletters}{OT1}{cmss}{m}{n}
\DeclareMathSymbol{\bsfGamma}{0}{bsfletters}{'000}
\DeclareMathSymbol{\ssfGamma}{0}{ssfletters}{'000}
\DeclareMathSymbol{\bsfDelta}{0}{bsfletters}{'001}
\DeclareMathSymbol{\ssfDelta}{0}{ssfletters}{'001}
\DeclareMathSymbol{\bsfTheta}{0}{bsfletters}{'002}
\DeclareMathSymbol{\ssfTheta}{0}{ssfletters}{'002}
\DeclareMathSymbol{\bsfLambda}{0}{bsfletters}{'003}
\DeclareMathSymbol{\ssfLambda}{0}{ssfletters}{'003}
\DeclareMathSymbol{\bsfXi}{0}{bsfletters}{'004}
\DeclareMathSymbol{\ssfXi}{0}{ssfletters}{'004}
\DeclareMathSymbol{\bsfPi}{0}{bsfletters}{'005}
\DeclareMathSymbol{\ssfPi}{0}{ssfletters}{'005}
\DeclareMathSymbol{\bsfSigma}{0}{bsfletters}{'006}
\DeclareMathSymbol{\ssfSigma}{0}{ssfletters}{'006}
\DeclareMathSymbol{\bsfUpsilon}{0}{bsfletters}{'007}
\DeclareMathSymbol{\ssfUpsilon}{0}{ssfletters}{'007}
\DeclareMathSymbol{\bsfPhi}{0}{bsfletters}{'010}
\DeclareMathSymbol{\ssfPhi}{0}{ssfletters}{'010}
\DeclareMathSymbol{\bsfPsi}{0}{bsfletters}{'011}
\DeclareMathSymbol{\ssfPsi}{0}{ssfletters}{'011}
\DeclareMathSymbol{\bsfOmega}{0}{bsfletters}{'012}
\DeclareMathSymbol{\ssfOmega}{0}{ssfletters}{'012}

\renewcommand{\defeq}{\triangleq}

\newcommand{\rvb}{{\mathssf{b}}}	% b
\newcommand{\rvbH}{{\mathbsf{H}}}

\newcommand{\rvt}{{\mathssf{t}}}	% t

\newcommand{\rvu}{{\mathssf{u}}}	% u

\newcommand{\rvbu}{{\mathbsf{u}}}

\newcommand{\rvbv}{{\mathbsf{v}}}

\newcommand{\rvw}{{\mathssf{w}}}	% w

\newcommand{\rvbw}{{\mathbsf{w}}}

\newcommand{\rvx}{{\mathssf{x}}}	% x, random variable

\newcommand{\rvbx}{{\mathbsf{x}}}

\newcommand{\rvy}{{\mathssf{y}}}	% y

\newcommand{\rvby}{{\mathbsf{y}}}
\newcommand{\rvbyt}{{\tilde{\rvby}}}

\newcommand{\rvbz}{{\mathbsf{z}}}
\newcommand{\rvbzt}{{\tilde{\rvbz}}}